\documentclass[pra,11pt,onecolumn,notitlepage]{revtex4-2}

\usepackage{amsmath}
\usepackage{mathtools}
\usepackage{multirow}
\usepackage{bbm, dsfont}
\usepackage{braket}
\usepackage{caption}
\usepackage{subcaption}
\usepackage{xcolor}
\usepackage{float}
\allowdisplaybreaks
\usepackage[utf8]{inputenc}
\usepackage{graphicx}
\newcommand{\RN}[1]{%
  \textup{\uppercase\expandafter{\romannumeral#1}}%
}
\begin{document}
	\title{Bidirectional quantum teleportation using quantum walks}
	\author{ A. S. Abay Krishna}
	\affiliation{%
			Department of Physics, Government Arts and Science College, Meenchanda, Calicut, Kerala 673018, India
	}%
	\author{K. K. Naseeda}
	\affiliation{%
			Department of Physics, Government  College, Malappuram, Kerala 676509, India
	}%
        \author{ N. C. Randeep}
	\affiliation{%
			Department of Physics, Government Arts and Science College, Meenchanda, Calicut, Kerala 673018, India
	}%
	\date{\today}

\begin{abstract}
We present a method for bidirectional teleportation of a single qubit using quantum walks on two independent one-dimensional lattices and two independent cycles with four vertices, employing nearest-neighbor jumps with coin outcomes. In addition, we discuss two different methods for two-qubit teleportation by employing nearest-neighbor jumps and next-nearest-neighbor jumps with a single coin and two coins, respectively. Finally, it is demonstrated that the two-qubit single-jump and the two-jump quantum walk teleportation schemes yield the same results.
\end{abstract}
\maketitle
\section{Introduction \label{sec1}}
 Quantum teleportation is a process where an unknown qubit state is transferred from one person (Alice) to another who is spatially separated (Bob)\cite{BEN1993}. This transfer is achieved by sharing an entangled state and utilizing classical communication between Alice and Bob. The first teleportation scheme was established by Bennett et al. in $1993$ and was later realized experimentally \cite{BEN1993,BOU1997}. Subsequently, quantum teleportation schemes for multi-qubit systems and higher-dimensional systems have been developed \cite{CHA2008,LUO2019,AMR2010,SEB2023,RAN2024}. Instead of a unidirectional teleportation scheme, researchers have developed a bidirectional teleportation scheme, where Alice and Bob simultaneously teleport qubits to each other \cite{Zha2013,Yan2013}. 
 
Although entanglement is crucial in this process, there is an alternative teleportation technique where initial entanglement between states is not required; instead, it is generated during the step-by-step teleportation process \cite{WAN2017}.  This can be achieved through a process known as a quantum walk \cite{VEN2012}, which is the quantum analogue of a classical walk, particularly through the use of a coined quantum walk \cite{AMB2001}. Quantum walks play an essential role in quantum computing, quantum simulation, quantum information processing, and graph-theoretic applications, and have been regarded as a promising computational model in recent years \cite{KAD2021,QIA2024}. Using coined quantum walk many protocols for teleportation of qubits have been developed \cite{LI2019,SHI2022,ZAR2023,Sha2018,Cha2019,Li2022,Yam2021,Che2022}. But bidirectional teleportation of qubits is not explored yet.

In this work, we investigate bidirectional quantum teleportation using the quantum walk technique. First, a one-dimensional quantum walk on two independent one-dimensional lattices is explored, with one lattice assigned to Alice and another to Bob.  Depending on the outcomes of their respective coins, Alice and Bob perform nearest-neighbor jumps on the lattice and execute single-qubit bidirectional teleportation. The discussion also includes single-qubit bidirectional teleportation on two independent $4$-cycles (one-dimensional periodic lattices with $4$ sites).  
Furthermore, two approaches to two-qubit bidirectional teleportation are presented: the first method involves a quantum walk on four independent one-dimensional lattices with nearest-neighbor jumps based on the outcomes of a single coin, and the second method involves a quantum walk on two independent lattices with both nearest-neighbor and next-nearest-neighbor jumps based on the outcomes of two coins. Finally, we conclude that in these two teleportation schemes, a mapping can be found between the position basis of the lattices, and in each equivalent case, the same unitary operators can reproduce the teleported state. 

The paper is organized as follows. In Sec.~\ref{sec2}, we discuss bidirectional quantum teleportation of a single qubit state. Sec.\ref{sec3}, presents a study of bidirectional teleportation using a quantum walk on a $4$-cycle. Sec.~\ref{sec4}, explains two-qubit bidirectional teleportation using a single-step quantum walk, while Sec.~\ref{sec5} proposes an alternative method based on a two-step quantum walk and compares it with the former. Finally we conclude in Sec.~\ref{sec6}.

\section{Bidirectional quantum teleportation of single qubit state \label{sec2}}
\begin{figure}
    \begin{center}
    \includegraphics[scale=1.3]{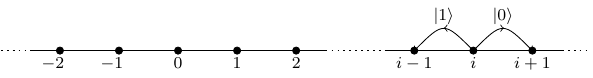}
    \end{center}
    \caption{The quantum walk on a one dimensional line. If the coin toss results in $0$, the walker moves from position $i$ to $i+1$ (rightward). If the coin toss result in $1$, the walker moves from $i$ to $i-1$ (leftward).}
    \label{Figure1}
    \end{figure}

In the bidirectional quantum walk process, Alice aims to teleport an unknown state to Bob, while Bob simultaneously aims to teleport an unknown state to Alice. This is accomplished by Alice and Bob each having three particles: the first particle of each holds the position space, while the second and third particles hold the coin spaces. The conditional shift operator can generate entanglement between the position space and the coin space. The conditional shift operators for this teleportation are defined as follows: 
\begin{equation}
\begin{split}
  S_{0}=&\sum_{i} |i+1\rangle\langle i|, \\
 S_{0}^\dag=&S_{1}=\sum_{i}|i\rangle \langle i+1|. 
  \label{Eq1}
\end{split}
\end{equation}
where $S_{0}$ and $S_{0}^{\dagger}=S_{1}$ represent the right shifting and left shifting operators, respectively, as illustrated in FIG.~(\ref{Figure1}). When the coin outcome is $0$, the operator $S_0$ moves the particle to its nearest neighboring site to the right (from site $i$ to $i+1$). Conversely, when the coin outcome is $1$, the operator $S_{1}$ moves the particle to its nearest neighboring site to the left  (from site $i$ to $i-1$).
    
 Suppose Alice wants to transmit an unknown state $|\phi_{1}\rangle=a_{0}|0\rangle + a_{1}|1\rangle$ to Bob, and Bob wants to transmit an unknown state $|\phi_{2}\rangle=b_0|0\rangle+b_1|1\rangle$ to Alice. This can be achieved simultaneously using quantum walk techniques. For this purpose, Alice has three particles: $A_{1},A_{2}$ and $A_{3}$, where the state of the first particle, $A_{1}$, represents the  position space  $P_{A_{1}}$,  and the states of $A_{2}$ and $A_{3}$ represent coin spaces $C_{1}$ and $C_{2}$, respectively. Similarly, Bob has three particles: $B_{1},B_{2}$ and $B_{3}$, where the state of $B_{1}$ represents the position space $P_{B_{1}}$, and the states of $B_{2},B_{3}$ represent the coin spaces $C_{3}$ and $C_{4}$, respectively.

 Initially, the unknown state that Alice wants to teleport is encoded in coin $C_{1}$, and the unknown state that Bob wants to teleport is encoded in coin $C_{3}$. Additionally, the position spaces $P_{A_{1}}$ and $P_{B_{1}}$ are initialized to $|0\rangle$, and the states of coins $C_2$ and $C_4$ are set to  $|0\rangle$. Thus, the total initial state, arranged in a convenient order, is
  \begin{equation}
|\psi_0\rangle_{A_{1}B_{1}A_{2}A_{3}B_{2}B_{3}}=|00\rangle\otimes(a_0|0\rangle+a_1|1\rangle)\otimes|0\rangle\otimes(b_0|0\rangle+b_1|1\rangle)\otimes|0\rangle.
    \label{Eq2}
 \end{equation}
The bidirectional quantum teleportation process involves four distinct walks, which are as follows:
\begin{equation}
 \begin{split}
     W_1 =&E_{1}(I_{A_{1}}\otimes I_{B_{1}}\otimes I_{A_{2}}\otimes I_{A_{3}}\otimes I_{B_{2}}\otimes I_{B_{3}}), \\
     W_{2}=&E_{2}(I_{A_{1}}\otimes I_{B_{1}}\otimes I_{A_{2}}\otimes I_{A_{3}}\otimes I_{B_{2}}\otimes I_{B_{3}}),  \\
     W_{3}=&E_3(I_{A_{1}}\otimes I_{B_{1}}\otimes I_{A_{2}}\otimes H_{A_{3}}\otimes I_{B_{2}}\otimes I_{B_{3}}),\\
    W_{4}=&E_{4}(I_{A_{1}}\otimes I_{B_{1}}\otimes I_{A_{2}}\otimes I_{A_{3}}\otimes I_{B_{2}}\otimes H_{B_{3}}),
     \end{split}
     \end{equation}
where  
\begin{equation}
 \begin{split}
E_1=&\sum\limits_{i=0}^{1}(S_{i}\otimes I_{B_{1}}\otimes|i\rangle\langle i|  \otimes I_{A_{3}} \otimes I_{B_{2}} \otimes I_{B_{3}}), \\
E_{2}=&\sum\limits_{i=0}^{1}( I_{A_{1}} \otimes S_{i} \otimes I_{A_{2}}\otimes I_{A_{3}} \otimes|i\rangle\langle i|\otimes I_{B_{3}}), \\
E_{3}=&\sum\limits_{i=0}^{1}( I_{A_1} \otimes S_{i} \otimes I_{A_2} \otimes|i\rangle\langle i|\otimes I_{B_2}\otimes I_{B_3}), \\
E_4=&\sum\limits_{i=0}^{1}(S_{i}\otimes I_{B_1}\otimes I_{A_2}\otimes I_{A_3}\otimes I_{B_2}\otimes|i\rangle\langle i|).
\end{split}
\end{equation}   
\begin{figure}
    \begin{center}
    \includegraphics[scale=.8]{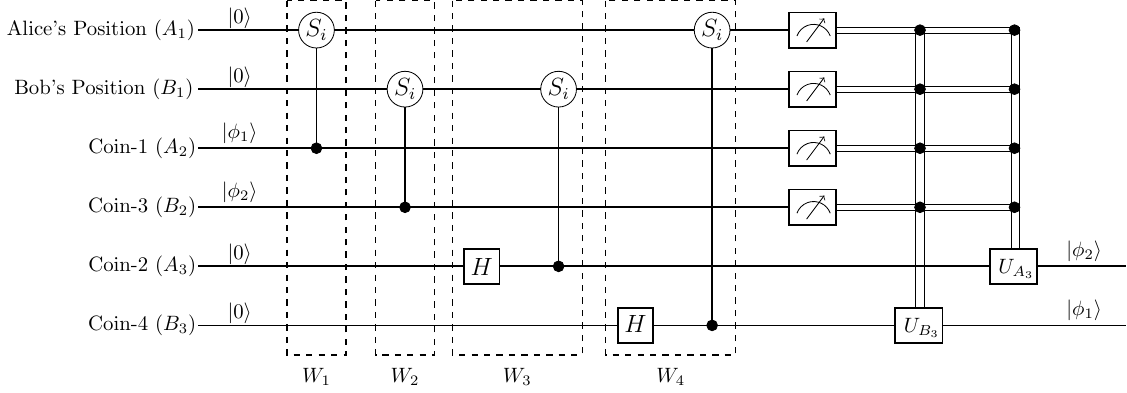}
    \end{center}
    \caption{Circuit diagram for bidirectional single qubit teleportation using quantum walk.}
    \label{Figure2}
    \end{figure}

In the above definitions of quantum walks, $I_{A_{i}}$   and $I_{B_{i}}$ are identity operators acting on the respective position/coin spaces of Alice and Bob. Similarly, $H_{A_{i}}$ and $H_{B_{i}}$  are Hadamard operators acting in the respective bases.

Now, we can analyze the protocol in detail by first applying the initial step of the quantum walk, $W_{1}$, to Eq. (\ref{Eq2}),
\begin{equation}
 \begin{split}
|\psi_1\rangle=a_{0}b_{0}|100000\rangle+a_{0}b_{1}|100010\rangle+a_{1}b_{0}|-101000\rangle+a_{1}b_{1}|-101010\rangle.
\end{split}
\end{equation}
Here, $|100000 \rangle=|100000 \rangle_{A_{1}B_{1}A_{2}A_{3}B_{2}B_{3}}$, represents the combined state of Alice and Bob, where $A_{1}$ and $B_{1}$ denote their one-dimensional position spaces, and 
$A_{2}, A_{3}, B_{2}, B_{3}$  correspond to their respective coin spaces. In the following calculation, a term like $|1-1 \rangle$ may arise. Since these represent different positions, they cannot be added to obtain the result $|0\rangle$.

Next, the second step, $W_2$, yields the following state
\begin{equation}
     \begin{split}
        |\psi_2\rangle=a_{0}b_{0}|110000\rangle+a_{0}b_{1}|1-10010\rangle+a_{1}b_{0}|-111000\rangle+a_{1}b_{1}|-1-11010\rangle.
     \end{split}
 \end{equation}
 Next, applying the third step of the quantum walk, $W_{3}$, transforms the state above to
 \begin{equation}
 \begin{split}
     |\psi_3\rangle= &\frac{1}{\sqrt{2}}\big(a_0b_0|120000\rangle+a_0b_0|100100\rangle+a_0b_1|100010\rangle+a_0b_1|1-20110\rangle+
     \\ &a_1b_0|-121000\rangle+a_1b_0|-101100\rangle+a_1b_1|-101010\rangle+a_1b_1|-1-21110\rangle\big)
 \end{split}
 \end{equation}
 Finally, the application of $W_{4}$ yields the final state
 \begin{equation}
\begin{split}
 |\psi_4\rangle=&\frac{1}{2}\big(a_{0}b_{0}|220000\rangle +a_{0}b_{0}|020001\rangle+ a_{0}b_{0}|200100\rangle+ a_{0}b_{0}|000101\rangle+a_{0}b_{1}|200010\rangle+ \\ &a_{0}b_{1}|000011\rangle+  a_{0}b_{1}|2-20110\rangle+a_{0}b_{1}|0-20111\rangle+a_{1}b_{0}|021000\rangle+ \\ & a_{1}b_{0}|-221001\rangle+a_{1}b_{0}|001100\rangle+ a_{1}b_{0}|-201101\rangle + a_{1}b_{1}|001010\rangle+ \\ &a_{1}b_{1}|-201011\rangle+ a_{1}b_{1}|0-21110\rangle+ a_{1}b_{1}|-2-21111\rangle\big).
 \label{Eq8}
 \end{split}
\end{equation}
To obtain the teleported state, we perform measurements across all position states and select components of the coin states. If the measurement results in both Alice's and Bob's position basis being $|00 \rangle$, then Eq.~(\ref{Eq8}) becomes
\begin{equation}
    |\psi_5\rangle_{A_{2}A_{3}B_{2}B_{3}}=\frac{1}{2}\big(a_{0}b_{0}|0101\rangle+a_{0}b_{1}|0011\rangle+ a_{1}b_{0} |1100\rangle+a_{1}b_{1}|1010\rangle\big).
\end{equation}
\noindent
Finally, particles $A_{2}$ and $B_{2}$ perform measurements  in the $|+\rangle |+\rangle$ basis, yielding
\begin{equation}
  |\psi_6\rangle_{A_{3}B_{3}}= \big(b_{0}|1\rangle +b_{1}|0\rangle\big)_{A_{3}}\big(a_{0}|1\rangle+a_{1}|0\rangle\big)_{B_{3}}
   \label{Eq10}
\end{equation}
with a probability of $\frac{1}{16}$. After the application of the unitary operations, $\sigma_{A_{3}}^{x}$ and $\sigma_{B_{3}}^{x}$ to the state described in Eq.~(\ref{Eq10}), the resulting state is obtained
\begin{equation}
  |\psi_7\rangle_{A_{3}B_{3}}= \big(b_{0}|0\rangle +b_{1}|1\rangle\big)_{A_{3}}\big(a_{0}|0\rangle+a_{1}|1\rangle\big)_{B_{3}}.
   \label{Eq11}
\end{equation}
That is, Alice and Bob can achieve bidirectional teleportation between each other. Instead of measuring in the $|+\rangle |+\rangle$, $A_{2}$ and $B_{2}$ can perform measurements in the $|+\rangle |-\rangle$,  $|-\rangle |+\rangle$ and $|-\rangle |-\rangle$ bases. The states intended for teleportation in both directions can be reconstructed through suitable unitary transformations, as illustrated in TABLE \ref{tab1}. Circuit diagram for above bidirectional bidirectional quantum teleportation is shown in FIG.~\ref{Figure2}. 

Similarly, Alice and Bob can perform measurements in the following bases: $ \frac{1}{\sqrt{2}}\big(|02\rangle\pm |0 -2\rangle\big),~ \frac{1}{\sqrt{2}}\big(|20\rangle \pm |-2 0\rangle\big),~ \frac{1}{2}\big(|22\rangle + |2 -2\rangle + |-2 2\rangle + |-2 -2\rangle \big),~ \frac{1}{2}\big(|22\rangle + |2 -2\rangle - |-2 2\rangle - |-2 -2\rangle \big),~ \frac{1}{2}\big(|22\rangle - |2 -2\rangle + |-2 2\rangle - |-2 -2\rangle \big),~ \frac{1}{2}\big(|22\rangle - |2 -2\rangle - |-2 2\rangle + |-2 -2\rangle \big)$, in the position basis of the particles $A_{1}$ and $B_{1}$ and $|+\rangle |+\rangle$, $|+\rangle |-\rangle$,  $|-\rangle |+\rangle$ and $|-\rangle |-\rangle$ in the coin basis of particles $A_{2}$ and $B_{2}$. All possible measurements and the corresponding unitary transformations for achieving bidirectional teleportation are presented in TABLE~\ref{tab1}. In the position bases $ \frac{1}{\sqrt{2}}\big(|02\rangle\pm |0 -2\rangle\big)$, and  $\frac{1}{\sqrt{2}}\big(|20\rangle\pm |-2 0\rangle\big)$,  after performing measurements in the coin basis of $A_{2}$ and $B_{2}$ and applying the unitary transformation as shown in TABLE~\ref{tab1}, in each case, the teleported state is reconstructed with a probability of 
 $\frac{1}{32}$. Similarly, in the four orthogonal position bases formed by the set $\{|2 2\rangle,~ | 2 -2 \rangle,~ |-2 2 \rangle,~ |-2 -2 \rangle \} $ as discussed above, after performing measurements and applying the unitary transformation as shown in TABLE~\ref{tab1}, we can reconstruct the teleported state with a probability of $\frac{1}{64}$ in each case.

\begin{table}
\caption{ Measurements performed by Alice and Bob in the position basis of particles $A_{1}$ and $B_{1}$ and in the coin basis of the particles $A_{2}$ and $B_{2}$, followed by the corresponding unitary transformations to reproduce the teleported states in bidirectional teleportation.}
    \centering
    \begin{tabular}{c c c}
    \hline
   Position basis $A_{1}$ and $B_{1}$ & ~~Coin basis of $A_{2}$  and $B_{2}$~&~ Unitary operations ~~ \\[1.ex]
    \hline
      & $\ket{+}_{A_{2}}\ket{+}_{B_{2}}$ & $\sigma_{A_{3}}^{x}\sigma_{B_{3}}^{x}$\\
& $\ket{+}_{A_{2}}\ket{-}_{B_{2}}$ & $\sigma_{A_{3}}^{z}\sigma_{A_{3}}^{x}\sigma_{B_{3}}^{x}$\\
$\ket{00}_{A_{1} B_{1}}$ & $\ket{-}_{A_{2}}\ket{+}_{B_{2}}$ & $\sigma_{B_{3}}^{z}\sigma_{A_{3}}^{x}\sigma_{B_{3}}^{x}$\\
& $\ket{-}_{A_{2}}\ket{-}_{B_{2}}$ & $\sigma_{A_{3}}^{z}\sigma_{B_{3}}^{z}\sigma_{A_{3}}^{x}\sigma_{B_{3}}^{x}$\\ 
\hline
 & $\ket{+}_{A_{2}}\ket{+}_{B_{2}}$ & $\sigma_{B_{3}}^{x}$\\
& $\ket{+}_{A_{2}}\ket{-}_{B_{2}}$ & $\sigma_{A_{3}}^{z}\sigma_{B_{3}}^{x}$\\
$\frac{1}{\sqrt{2}}(\ket{02}+\ket{0-2})_{A_{1}B_{1}}$ & $\ket{-}_{A_{2}}\ket{+}_{B_{2}}$ & $\sigma_{B_{3}}^{z}\sigma_{B_{3}}^{x}$\\
& $\ket{-}_{A_{2}}\ket{-}_{B_{2}}$ & $\sigma_{A_{3}}^{z}\sigma_{B_{3}}^{z}\sigma_{B_{3}}^{x}$\\
\hline
 & $\ket{+}_{A_{2}}\ket{+}_{B_{2}}$ & $\sigma_{B_{3}}^{z}\sigma_{B_{3}}^{x}$\\
& $\ket{+}_{A_{2}}\ket{-}_{B_{2}}$ & $\sigma_{B_{3}}^{x}$\\
$\frac{1}{\sqrt{2}}(\ket{02}-\ket{0-2})_{A_{1} B_{1}}$ & $\ket{-}_{A_{2}}\ket{+}_{B_{2}}$ & $\sigma_{B_{3}}^{z}\sigma_{A_{3}}^{z}\sigma_{B_{3}}^{x}$\\
& $\ket{-}_{A_{2}}\ket{-}_{B_{2}}$ & $\sigma_{B_{3}}^{z}\sigma_{B_{3}}^{x}$\\
\hline
 & $\ket{+}_{A_{2}}\ket{+}_{B_{2}}$ & $\sigma_{A_{3}}^{x}$\\
& $\ket{+}_{A_{2}}\ket{-}_{B_{2}}$ &  $\sigma_{A_{3}}^{z}\sigma_{A_{3}}^{x}$\\
$\frac{1}{\sqrt{2}}(\ket{20}+\ket{-20})_{A_{1} B_{1}}$ & $\ket{-}_{A_{2}}\ket{+}_{B_{2}}$ &  $\sigma_{B_{3}}^{z}\sigma_{A_{3}}^{x}$\\
& $\ket{-}_{A_{2}}\ket{-}_{B_{2}}$ &  $\sigma_{B_{3}}^{z}\sigma_{A_{3}}^{z}\sigma_{A_{3}}^{x}$\\
\hline
 & $\ket{+}_{A_{2}}\ket{+}_{B_{2}}$ & $\sigma_{B_{3}}^{z}\sigma_{A_{3}}^{x}$\\
& $\ket{+}_{A_{2}}\ket{-}_{B_{2}}$ & $\sigma_{B_{3}}^{z}\sigma_{A_{3}}^{z}\sigma_{A_{3}}^{x}$\\
$\frac{1}{\sqrt{2}}(\ket{20}-\ket{-20})_{A_{1}B_{1}}$ & $\ket{-}_{A_{2}}\ket{+}_{B_{2}}$ & $\sigma_{A_{3}}^{x}$\\
& $\ket{-}_{A_{2}}\ket{-}_{B_{2}}$ & $\sigma_{A_{3}}^{z}\sigma_{A_{3}}^{x}$\\ 
\hline
 & $\ket{+}_{A_{2}}\ket{+}_{B_{2}}$ & I\\
& $\ket{+}_{A_{2}}\ket{-}_{B_{2}}$ & $\sigma_{A_{3}}^{z}$\\
$\frac{1}{2}(\ket{22}+\ket{2-2}+\ket{-22}+\ket{-2-2})_{A_{1} B_{1}}$ & $\ket{-}_{A_{2}}\ket{+}_{B_{2}}$ & $\sigma_{B_{3}}^{z}$\\
& $\ket{-}_{A_{2}}\ket{-}_{B_{2}}$ & $\sigma_{B_{3}}^{z}\sigma_{A_{3}}^{z}$\\
\hline
 & $\ket{+}_{A_{2}}\ket{+}_{B_{2}}$ & $\sigma_{A_{3}}^{z}$\\
& $\ket{+}_{A_{2}}\ket{-}_{B_{2}}$ & I\\
$\frac{1}{2}(\ket{22}-\ket{2-2}+\ket{-22}-\ket{-2-2})_{A_{1} B_{1}}$ & $\ket{-}_{A_{2}}\ket{+}_{B_{2}}$ & $\sigma_{A_{3}}^{z}$$\sigma_{B_{3}}^{z}$\\
& $\ket{-}_{A_{2}}\ket{-}_{B_{2}}$ &$\sigma_{B_{3}}^{z}$\\
\hline
 & $\ket{+}_{A_{2}}\ket{+}_{B_{2}}$ & $\sigma_{B_{3}}^{z}$\\
& $\ket{+}_{A_{2}}\ket{-}_{B_{2}}$ & $\sigma_{B_{3}}^{z}$$\sigma_{A_{3}}^{z}$\\
$\frac{1}{2}(\ket{22}+\ket{2-2}-\ket{-22}-\ket{-2-2})_{A_{1} B_{1}}$ & $\ket{-}_{A_{2}}\ket{+}_{B_{2}}$ & I \\
& $\ket{-}_{A_{2}}\ket{-}_{B_{2}}$ & $\sigma_{A_{3}}^{z}$\\ 
\hline
 & $\ket{+}_{A_{2}}\ket{+}_{B_{2}}$ & $\sigma_{B_{3}}^{z}\sigma_{A_{3}}^{z}$\\
& $\ket{+}_{A_{2}}\ket{-}_{B_{2}}$ & $\sigma_{B_{3}}^{z}$\\
$\frac{1}{2}(\ket{22}-\ket{2-2}-\ket{-22}+\ket{-2-2})_{A_{1} B_{1}}$ & $\ket{-}_{A_{2}}\ket{+}_{B_{2}}$ & $\sigma_{A_{3}}^{z}$\\
& $\ket{-}_{A_{2}}\ket{-}_{B_{2}}$ & I\\
       [1ex]
       \hline
    \end{tabular}
    \label{tab1}
\end{table}

\section{Bidirectional Teleportation by quantum walk on 4-cycle \label{sec3}}
\begin{figure}
    \begin{center}
    \includegraphics[scale=.9]{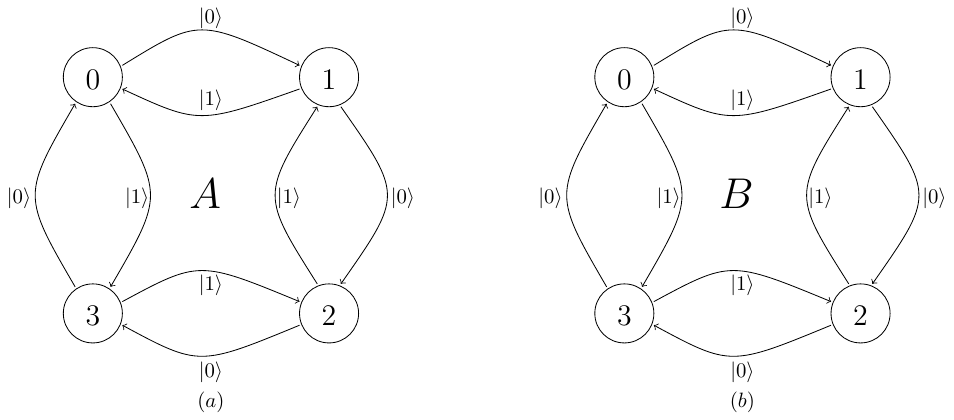}
    \end{center}
    \caption{Two independent cyclic position basis ($4$ cycle) on a one dimensional line, figure (a) for Alice and figure (b) for Bob.}
    \label{Figure3}
    \end{figure}

In this teleportation scheme, the position spaces of Alice and Bob are two independent $4$-cycles with vertices  $\ket{0},\ket{1},\ket{2}$ and $\ket{3}$ in each, as shown in FIG. \ref{Figure3}. As in the previous section, Alice has three particles: $A_{1}$, $A_{2}$, and $A_{3}$, where the state of  $A_{1}$ represents Alice's  position space ($4$-cycle), and the states of   $A_{2}$ and $A_{3}$ are coin spaces $C_{1}$ and $C_{2}$, respectively. Similarly, $B_{1}$ represents Bob's position space ($4$-cycle) and the states of $B_{2}$ and $B_{3}$ are coin spaces $C_{3}$ and $C_{4}$, respectively.
    
Initially, the unknown state to be teleported by Alice $(a_{0}\ket{0}+a_{1}\ket{1})$ is encoded in coin $C_{1}$, and the unknown state to be teleported by Bob $(b_{0}\ket{0}+b_{1}\ket{1})$ is encoded in the $C_{3}$. Additionally, initialize  the position spaces of $A_{1}$ and $B_{1}$ to $|0\rangle$ and coin spaces of $C_2$ and $C_4$ to $\ket{+}=\frac{1}{\sqrt{2}}(\ket{0}+\ket{1})$. Then the total initial state, in the convenient order, is   
\begin{equation}
    \begin{split}
\ket{\psi_{0}}_{A_{1}B_{1}A_{2}A_{3}B_{2}B_{3}} = \ket{0}\otimes \ket{0}\otimes (a_{0}\ket{0}+a_{1}\ket{1})\otimes \ket{+} \otimes(b_{0}\ket{0}+b_{1}\ket{1})\otimes \ket{+}
\end{split}
\label{Eqc1}
\end{equation}
The conditional shift operator in cyclic teleportation is defined as                                                  \begin{equation}
   S= \sum\limits_{i=0}^{2} |i+1\rangle \langle i| \otimes |0\rangle \langle 0 |+ \sum\limits_{i=0}^{3} |i-1\rangle \langle i| \otimes |1\rangle \langle 1 |+|0 \rangle \langle 3 | \otimes |0\rangle \langle 0 |+ |3 \rangle \langle 0 | \otimes |1\rangle \langle 1 |
\end{equation}
As in Sec.\ref{sec2}, after four steps of the quantum walk, Eqn.(\ref{Eqc1}) becomes
\begin{equation}
    \begin{split}
    \ket{\psi_1}=\frac{1}{2}(a_0b_0\ket{220000}+a_0b_0\ket{200100}+a_0b_0\ket{020001}+a_0b_0\ket{000101}\\+a_1b_0\ket{021000}+a_1b_0\ket{001100}+a_1b_0\ket{221001}+a_1b_0\ket{201101}\\+a_0b_1\ket{200010}+a_0b_1\ket{220110}+a_0b_1\ket{000011}+a_0b_1\ket{020111}\\+a_1b_1\ket{001010}+a_1b_1\ket{021110}+a_1b_1\ket{201011}+a_1b_1\ket{221111})
    \end{split}
\end{equation}
If Alice and Bob jointly measure $|00\rangle$ in their position space, the resulting state is given by
\begin{equation}
    \begin{split}
    \ket{\psi_{2}}=\frac{1}{2}(a_{0}b_{0}\ket{0101}+a_{0}b_{1}\ket{0011}+a_{1}b_{0}\ket{1100}+a_{1}b_{1}\ket{1111})
        \end{split}
\end{equation}
and then measure $\ket{+}$ on both of their coin spaces $A_{2}$  and $B_{2}$,  the final state will be
\begin{equation}
    \begin{split}
    \ket{\psi_{3}}=(b_{0}\ket{1}+b_{1}\ket{0})_{A_{2}}\otimes(a_{0}\ket{1}+a_{1}\ket{0})_{B_{2}}\\
    \end{split}
\end{equation}
Then, applying $\sigma_{x}$ operations to both $A_{2}$ and $B_{2}$ particles will produce 
\begin{equation}
    \begin{split}
    \ket{\psi_{4}}=(b_{0}\ket{0}+b_{1}\ket{1})_{A_{2}}\otimes(a_{0}\ket{0}+a_{1}\ket{1})_{B_{2}}.
    \end{split}
\end{equation}
That is, Alice's $A_{2}$ reproduces the state sent by Bob's $B_{1}$, and Bob's $B_{2}$ reproduces the state sent by Alice's $A_{1}$. Similarly, for the other position and coin bases, Alice $A_{2}$ and $B_{2}$ reproduce the states send by Bob's $B_{1}$ and Alice's $A_{1}$, respectively, as shown in Table.~\ref{tab2}.

\begin{table}[ht]
\caption{outcome of measurement performed by Alice and Bob in the first coin state and the unitary operation on their second coin state for reproducing the unknown single qubit state }
\centering
\begin{tabular}{c c c}
\hline
Position basis & ~~~~First coin basis $A_{2}$  and $B_{2}$~~~&~~~ Unitary operations ~~ \\[2.ex]
\hline
 & $\ket{+}_{A_{2}}\ket{+}_{B_{2}}$ & $\sigma_{A_{3}}^{x}\sigma_{B_{3}}^{x}$\\
$\ket{00}_{A_{1} B_{1}}$ & $\ket{+}_{A_{2}}\ket{-}_{B_{2}}$ & $\sigma_{A_{3}}^{z}\sigma_{A_{3}}^{x}\sigma_{B_{3}}^{x}$\\
 & $\ket{-}_{A_{2}}\ket{+}_{B_{2}}$ & $\sigma_{B_{3}}^{z}\sigma_{A_{3}}^{x}\sigma_{B_{3}}^{x}$\\
& $\ket{-}_{A_{2}}\ket{-}_{B_{2}}$ & $\sigma_{A_{3}}^{z}\sigma_{B_{3}}^{z}\sigma_{A_{3}}^{x}\sigma_{B_{3}}^{x}$ \\

\hline
 & $\ket{+}_{A_{2}}\ket{+}_{B_{2}}$ & $\sigma_{B_{3}}^{x}$\\
$\ket{02}_{A_{1} B_{1}}$ & $\ket{+}_{A_{2}}\ket{-}_{B_{2}}$ & $\sigma_{A_{3}}^{z}\sigma_{B_{3}}^{x}$\\
& $\ket{-}_{A_{2}}\ket{+}_{B_{2}}$ & $\sigma_{B_{3}}^{z}\sigma_{B_{3}}^{x}$\\
& $\ket{-}_{A_{2}}\ket{-}_{B_{2}}$ & $\sigma_{A_{3}}^{z}\sigma_{B_{3}}^{z}\sigma_{B_{3}}^{x}$\\ \hline

 & $\ket{+}_{A_{2}}\ket{+}_{B_{2}}$ & $\sigma_{A_{3}}^{x}$\\
$\ket{20}_{A_{1} B_{1}}$ & $\ket{+}_{A_{2}}\ket{-}_{B_{2}}$ &  $\sigma_{A_{3}}^{z}\sigma_{A_{3}}^{x}$\\
& $\ket{-}_{A_{2}}\ket{+}_{B_{2}}$ &  $\sigma_{B_{3}}^{z}\sigma_{A_{3}}^{x}$\\
& $\ket{-}_{A_{2}}\ket{-}_{B_{2}}$ &  $\sigma_{B_{3}}^{z}\sigma_{A_{3}}^{z}\sigma_{A_{3}}^{x}$\\ \hline
 & $\ket{+}_{A_{2}}\ket{+}_{B_{2}}$ & I\\
$\ket{22}_{A_{1} B_{1}}$ & $\ket{+}_{A_{2}}\ket{-}_{B_{2}}$ & $\sigma_{A_{3}}^{z}$\\
& $\ket{-}_{A_{2}}\ket{+}_{B_{2}}$ & $\sigma_{B_{3}}^{z}$\\
& $\ket{-}_{A_{2}}\ket{-}_{B_{2}}$ & $\sigma_{B_{3}}^{z}\sigma_{A_{3}}^{z}$\\ 
[1ex]
\hline
\end{tabular}
\label{tab2}
\end{table}

\section{BIDIRECTIONAL TELEPORTATION OF TWO QUBIT USING SINGLE STEP QUANTUM WALK \label{sec4}}
In this section, we discuss the extension of the bidirectional single-qubit teleportation described in Sec.\ref{sec2} to bidirectional two-qubit teleportation. Here, Alice and Bob each hold six particles: the first two particles represent their position spaces, and the remaining four particles represent their coin spaces. The conditional shift operator in this case is the same as in   Eq. (\ref{Eq1}) in Sec.\ref{sec2}.

In bidirectional two-qubit teleportation, the unknown state that Alice wants to send to Bob is $|\phi_{1}\rangle=a_{0}\ket{00}+a_{1}\ket{01}+a_{2}\ket{10}+a_{3}\ket{11}$, and the unknown state that Bob wants to send to Alice is $|\phi_{2}\rangle=b_{0}\ket{00}+b_{1}\ket{01}+b_{2}\ket{10}+b_{3}\ket{11}$. Let Alice have six particles: $A_{1}, A_{2}$,...$A_{6}$, where the states of the first two particles $A_{1}$ and $A_{2}$, represent the position spaces of Alice, $P_{A_{1}}$ and $P_{A_{2}}$, respectively. Next, the states of particles  $A_{3}, A_{4}, A_{5}$ and $A_{6}$ represent the coin spaces of the coins $C_{1}, C_{2}, C_{3}$ and $C_{4}$, respectively. Similarly, Bob also have six particles: $B_{1}, B_{2}$,...$B_{6}$, where the states of the first  particles, $B_{1}$ and $B_{2}$, represent Bob's position spaces,  $P_{B_{1}}$ and $P_{B_{2}}$, respectively. As in the case of Alice, states of particles  $B_{3}, B_{4}, B_{5}$ and $B_{6}$ represent coin spaces of the coins $C_{5}, C_{6}, C_{7}$ and $C_{8}$, respectively.

Initially, the unknown two state to be teleported by Alice is encoded in particles $A_{3}$ and $A_{4}$, and the unknown state to be teleported by Bob is encoded  in particles $B_{3}$ and $B_{4}$. Then, the position spaces of $A_{1}A_{2}$ and $B_{1}B_{2}$  initialized to $|00\rangle$, and coin spaces of $A_{5}, A_{6}$ and $B_{5}, B_{6}$ are initialized to $|00\rangle$. Thus, the total initial state in the convenient order is 
\begin{equation}
  \begin{split} |\psi_0\rangle_{A_{1}A_{2}B_{1}B_{2}A_{3}A_{4}A_{5}A_{6}B_{3}B_{4}B_{5}B_{6}}=&|0000\rangle\otimes(a_{0}|00\rangle + a_{1}|01\rangle + a_{2}|10\rangle + a_{3}|11\rangle)\otimes|00\rangle\otimes\\&(b_{0}|00\rangle + b_{1}|01\rangle + b_{2}|10\rangle + b_{3}|11\rangle)\otimes|00\rangle.
  \end{split}
  \label{Eqn5-1}
    \end{equation}
The bidirectional quantum teleportation process consists four walks, which are  
 \begin{equation}
 \begin{split}
     W_1 =&E_{1}(I_{A_{1}A_{2}}\otimes I_{B_{1}B_{2}}\otimes I_{A_{3}A_{4}}\otimes I_{A_{5}A_{6}}\otimes I_{B_{3}B_{4}}\otimes I_{B_{5}B_{6}}),\\ W_{2}=&E_{2} (I_{A_{1}A_{2}}\otimes I_{B_{1}B_{2}}\otimes I_{A_{3}A_{4}}\otimes I_{A_{5}A_{6}}\otimes I_{B_{3}B_{4}}\otimes I_{B_{5}B_{6}}) \\W_3=&E_3 (I_{A_{1}A_{2}}\otimes I_{B_{1}B_{2}}\otimes I_{A_{3}A_{4}}\otimes (H \otimes H)_{A_{5}H_{6}}\otimes I_{B_{3}B_{4}}\otimes I_{B_{5}B_{6}}),\\ W_{4}=&E_{4} (I_{A_{1}A_{2}}\otimes I_{B_{1}B_{2}}\otimes I_{A_{3}A_{4}}\otimes I_{A_{5}A_{6}} \otimes I_{B_{3}B_{4}}\otimes (H \otimes H)_{B_{5}B_{6}}),
 \end{split}
 \label{Eqn5-2}
  \end{equation} 
 where  
\begin{equation}
 \begin{split}
E_1=&\sum\limits_{i,j=0}^{1}(S_{i}\otimes S_{j}\otimes I_{B_{1} B_{2}}\otimes|ij\rangle\langle ij|  \otimes I_{A_{5}A_{6}} \otimes I_{B_{3}B_{4}} \otimes I_{B_{5}B_{6}}),\\
E_2=&\sum\limits_{i,j=0}^{1}(I_{A_{1}A_{2}} \otimes S_{i}\otimes S_{j}\otimes I_{A_{3} A_{4}}\otimes I_{A_{5}A_{6}} \otimes|ij\rangle\langle ij|   \otimes I_{B_{5}B_{6}}),\\
E_3=&\sum\limits_{i,j=0}^{1}(I_{A_{1}A_{2}} \otimes S_{i}\otimes S_{j}\otimes \otimes I_{A_{3}A_{4}} \otimes|ij\rangle\langle ij|  \otimes  I_{B_{3}B_{4}} \otimes I_{B_{5}B_{6}}),\\
E_4=&\sum\limits_{i,j=0}^{1}(S_{i}\otimes S_{j}\otimes I_{B_{1} B_{2}} \otimes I_{A_{3}A_{4}} \otimes I_{A_{5}A_{6}} \otimes I_{B_{3}B_{4}}\otimes|ij\rangle\langle ij| ).\\
 \end{split}
 \label{Eqn5-3}
  \end{equation}  
In the above expression, the notation $I_{A_{1}A_{2}}$
is used in place of $I_{A_{1}} \otimes I_{A_{2}}$.
After performing the four different quantum walks defined in Eqs.~(\ref{Eqn5-2}), (\ref{Eqn5-3}), (\ref{Eq1}) on the initial state given in Eq. (\ref{Eqn5-1}), the resulting state corresponds to that given in \textbf{Appendix A}, Eq.~(\ref{Eqn5final}).

\begin{figure}
    \begin{center}
    \includegraphics[scale=.65]{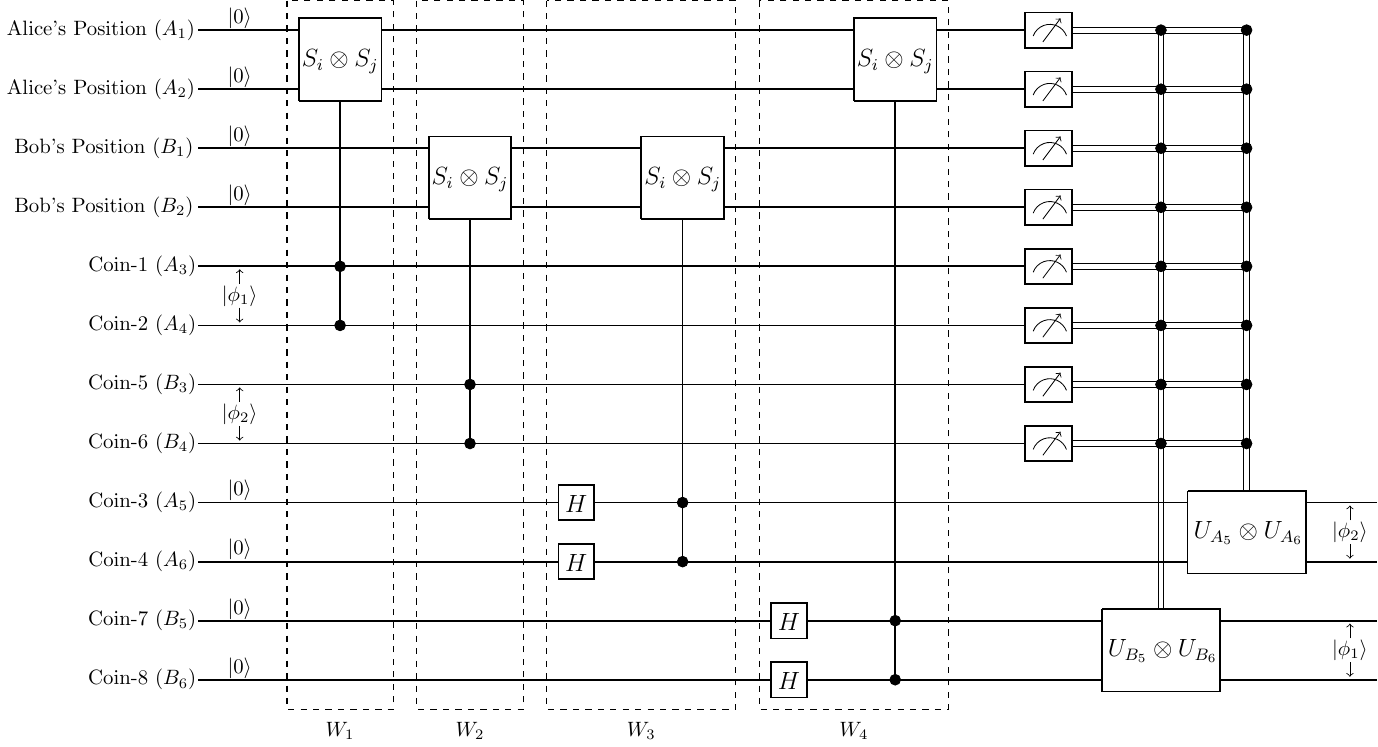}
    \end{center}
    \caption{Circuit diagram for bidirectional teleportation of two-qubit states using single step quantum walk.}
    \label{Figure4}
    \end{figure}

If Alice and Bob jointly perform a measurement on the final state in  Eqn.~(\ref{Eqn5final}) in the position basis in the state  $|0000\rangle_{A_{1}A_{2}B_{1}B_{2}}$, then the resulting state is given by  
\begin{equation}
    \begin{split}
    \ket{\psi_{2}}_{A_{3}A_{4}A_{5}A_{6}B_{3}B_{4}B_{5}B_{6}}=&\frac{1}{4}(a_{0}b_{0}\ket{00110011}+a_{0}b_{1}\ket{00100111}+a_{0}b_{2}\ket{00011011}+a_{0}b_{3}\ket{00001111}+\\&a_{1}b_{0}\ket{01110010}+a_{1}b_{1}\ket{01100110}+a_{1}b_{2}\ket{01011010}+a_{1}b_{3}\ket{01001110}+\\&a_{2}b_{0}\ket{10110001}+a_{2}b_{1}\ket{10100101}+a_{2}b_{2}\ket{10011001}+a_{2}b_{3}\ket{10001101}+\\&a_{3}b_{0}\ket{11110000}+a_{3}b_{1}\ket{11100100}+a_{3}b_{2}\ket{11011000}+a_{3}b_{3}\ket{11001100}).
        \end{split}
\end{equation}
Then, Alice and Bob perform measurements in the coin basis $|++\rangle_{A_{3}A_{4}}$ and $|++\rangle_{B_{3}B_{4}}$, respectively, resulting in the following state
\begin{equation}
    \begin{split}
    \ket{\psi_{3}}_{A_{5}A_{6}B_{5}B_{6}}=&\frac{1}{16}(a_{0}b_{0}\ket{1111}+a_{0}b_{1}\ket{1011}+a_{0}b_{2}\ket{0111}+a_{0}b_{3}\ket{0011}+a_{1}b_{0}\ket{1110}+a_{1}b_{1}\ket{1010}+\\&a_{1}b_{2}\ket{0110}+a_{1}b_{3}\ket{0010}+a_{2}b_{0}\ket{1101}+a_{2}b_{1}\ket{1001}+a_{2}b_{2}\ket{0101}+a_{2}b_{3}\ket{0001}+\\&a_{3}b_{0}\ket{1100}+a_{3}b_{1}\ket{1000}+a_{3}b_{2}\ket{0100}+a_{3}b_{3}\ket{0000}),
        \end{split}
\end{equation}
which can be written as
\begin{equation}
    \begin{split}
    \ket{\psi_{4}}=\frac{1}{16}[(b_{0}\ket{11}+b_{1}\ket{10}+b_{2}\ket{01}+b_{3}\ket{00})_{A_{5}A_{6}}\otimes(a_{0}\ket{11}+a_{1}\ket{10}+a_{2}\ket{01}+a_{3}\ket{00})_{B_{5}B_{6}}]\\
    \end{split}
\end{equation}
Then, applying the operations $(\sigma_{x})_{A_{4}}(\sigma_{x})_{A_{5}}(\sigma_{x})_{B_{4}}(\sigma_{x})_{B_{5}}$ on the above state, it becomes
\begin{equation}
    \begin{split}
    \ket{\psi_{5}}=\frac{1}{16}[(b_{0}\ket{00}+b_{1}\ket{01}+b_{2}\ket{10}+b_{3}\ket{11})_{A_{5}A_{6}}\otimes(a_{0}\ket{00}+a_{1}\ket{01}+a_{2}\ket{10}+a_{3}\ket{11})_{B_{5}B_{6}}]\\
    \end{split}
\end{equation}
That is, Alice is able to teleport an unknown state to Bob and Bob is able to teleport an unknown state
to Alice simultaneously. Table.~\ref{tab3} shows the remaining measurements on both Alice’s and Bob’s
coins, along with the corresponding unitary operations for bidirectional teleportation, when the position space measurement is $|0000\rangle_{A_{1}A_{2}B_{1}B_{2}}$. The detailed procedure for two qubit bidirectional quantum teleportation using a single step quantum walk is shown in FIG.~\ref{Figure4}.

\begin{table}
\caption{Outcomes of the measurements performed by Alice and Bob in their position basis $|0000\rangle$, and the first two coin bases of both, along with the corresponding unitary operations on their last two coins, for reproducing the unknown two-qubit states.}
\begin{tabular}{c  c c}
\hline
position basis~~~~~~~~  & ~~~~ coin basis  ~~~~&~ Unitary operation ~~ \\[1.ex]
\hline
 & $\ket{++}_{A_{3}A_{4}}\ket{++}_{B_{3}B_{4}}$ & $\sigma_{A_{5}}^{x}\sigma_{A_{6}}^{x}\sigma_{B_{5}}^{x}\sigma_{B_{6}}^{x}$  \\
& $\ket{++}_{A_{3}A_{4}}\ket{+-}_{B_{3}B_{4}}$ & $\sigma_{A_{5}}^{z}\sigma_{A_{5}}^{x}\sigma_{A_{6}}^{x}\sigma_{B_{5}}^{x}\sigma_{B_{6}}^{x}$ \\
& $\ket{++}_{A_{3}A_{4}}\ket{-+}_{B_{3}B_{4}}$ & $\sigma_{A_{5}}^{z}\sigma_{A_{5}}^{x}\sigma_{A_{6}}^{x}\sigma_{B_{5}}^{x}\sigma_{B_{6}}^{x}$ \\
& $\ket{++}_{A_{3}A_{4}}\ket{--}_{B_{3}B_{4}}$ & $\sigma_{A_{6}}^{z}\sigma_{A_{5}}^{z}\sigma_{A_{5}}^{x}\sigma_{A_{6}}^{x}\sigma_{B_{5}}^{x}\sigma_{B_{6}}^{x}$ \\
& $\ket{+-}_{A_{3}A_{4}}$$\ket{++}_{B_{3}B_{4}}$ &$\sigma_{A_{5}}^{x}\sigma_{A_{6}}^{x}\sigma_{B_{6}}^{z}\sigma_{B_{5}}^{x}\sigma_{B_{6}}^{x}$ \\
& $\ket{+-}_{A_{3}A_{4}}\ket{+-}_{B_{3}B_{4}}$ & $\sigma_{A_{6}}^{z}\sigma_{A_{5}}^{x}\sigma_{A_{6}}^{x}\sigma_{B_{6}}^{z}\sigma_{B_{5}}^{x}\sigma_{B_{6}}^{x}$ \\
& $\ket{+-}_{A_{3}A_{4}}\ket{-+}_{B_{3}B_{4}}$ & $\sigma_{A_{5}}^{z}\sigma_{A_{5}}^{x}\sigma_{A_{6}}^{x}\sigma_{B_{6}}^{z}\sigma_{B_{5}}^{x}\sigma_{B_{6}}^{x}$ \\
$\ket{0000}_{A_{1}A_{2}B_{1}B_{2}}$ & $\ket{+-}_{A_{3}A_{4}}\ket{--}_{B_{3}B_{4}}$ & $\sigma_{A_{6}}^{z}\sigma_{A_{5}}^{z}\sigma_{A_{5}}^{x}\sigma_{A_{6}}^{x}\sigma_{B_{6}}^{z}\sigma_{B_{5}}^{x}\sigma_{B_{6}}^{x}$ \\
& $\ket{-+}_{A_{3}A_{4}}\ket{++}_{B_{3}B_{4}}$ & $\sigma_{A_{5}}^{x}\sigma_{A_{6}}^{x}\sigma_{B_{5}}^{z}\sigma_{B_{5}}^{x}\sigma_{B_{6}}^{x}$\\
& $\ket{-+}_{A_{3}A_{4}}\ket{+-}_{B_{3}B_{4}}$ & $\sigma_{A_{6}}^{z}\sigma_{A_{5}}^{x}\sigma_{A_{6}}^{x}\sigma_{B_{5}}^{z}\sigma_{B_{5}}^{x}\sigma_{B_{6}}^{x}$ \\
& $\ket{-+}_{A_{3}A_{4}}\ket{-+}_{B_{3}B_{4}}$ & $\sigma_{A_{5}}^{z}\sigma_{A_{5}}^{x}\sigma_{A_{6}}^{x}\sigma_{B_{5}}^{z}\sigma_{B_{5}}^{x}\sigma_{B_{6}}^{x}$\\
& $\ket{-+}_{A_{3}A_{4}}\ket{--}_{B_{3}B_{4}}$ & $\sigma_{A_{6}}^{z}\sigma_{A_{5}}^{z}\sigma_{A_{5}}^{x}\sigma_{A_{6}}^{x}\sigma_{B_{5}}^{z}\sigma_{B_{5}}^{x}\sigma_{B_{6}}^{x}$\\
& $\ket{--}_{A_{3}A_{4}}\ket{++}_{B_{3}B_{4}}$ & $\sigma_{A_{5}}^{x}\sigma_{A_{6}}^{x}\sigma_{B_{6}}^{z}\sigma_{B_{5}}^{z}\sigma_{B_{5}}^{x}\sigma_{B_{6}}^{x} $ \\
& $\ket{--}_{A_{3}A_{4}}\ket{-+}_{B_{3}B_{4}}$ & $\sigma_{A_{5}}^{z}\sigma_{A_{5}}^{x}\sigma_{A_{6}}^{x}\sigma_{B_{6}}^{z}\sigma_{B_{5}}^{z}\sigma_{B_{5}}^{x}\sigma_{B_{6}}^{x}$\\
& $\ket{--}_{A_{3}A_{4}}\ket{--}_{B_{3}B_{4}}$ & $\sigma_{A_{6}}^{z}\sigma_{A_{5}}^{z}\sigma_{A_{5}}^{x}\sigma_{A_{6}}^{x}\sigma_{B_{6}}^{z}\sigma_{B_{5}}^{z}\sigma_{A_{5}}^{x}\sigma_{B_{6}}^{x}$ \\
[1.ex]
\hline
\end{tabular}
\label{tab3}
\end{table}

Similarly, measurements can be performed using the orthonormal states formed from the following disjoint set of states: $P_{1}=\{|0200\rangle, \ket{0-200}\},~ P_{2}=\{\ket{2000}, \ket{-2000} \},~P_{3}=\{ \ket{2200}, \ket{2-200},\\ \ket{-2200}, \ket{-2-200} \},~P_{4}=\{ \ket{0002}, \ket{000-2} \},~ P_{5}=\{\ket{0202}, \ket{020-2}, \ket{0-202}, \ket{0-20-2} \},\\~P_{6}=\{\ket{2002}, \ket{200-2}, \ket{-2002}, \ket{-200-2} \},~P_{7}=\{ \ket{2202}, \ket{220-2}, \ket{2-202}, \ket{2-20-2},\\ \ket{-2202}, \ket{-220-2}, \ket{-2-202},\ket{-2-20-2} \},~P_{8}=\{\ket{0020},\ket{00-20} \},~P_{9}=\{\ket{0220}, \ket{02-20},\\ \ket{0-220}, \ket{0-2-20} \},~P_{10}=\{\ket{2020}, \ket{20-20}, \ket{-2020}, \ket{-20-20} \},~P_{11}=\{\ket{2220},\ket{22-20},\\ \ket{2-220}, \ket{2-2-20}, \ket{-2220}, \ket{-22-20}, \ket{-2-220}, \ket{-2-2-20} \},~P_{12}=\{\ket{0022}, \ket{002-2},\\ \ket{00-22}, \ket{00-2-2} \},~P_{13}=\{\ket{0222}, \ket{022-2}, \ket{02-22}, \ket{02-2-2}, \ket{0-222}, \ket{0-22-2},\\ \ket{0-2-22},\ket{0-2-2-2} \},~P_{14}=\{\ket{2022}, \ket{202-2}, \ket{20-22}, \ket{20-2-2}, \ket{-2022}, \ket{-202-2},\\ \ket{-20-22}, \ket{-20-2-2} \},~P_{15}=\{\ket{2222}, \ket{222-2}, \ket{22-22}, \ket{22-2-2}, (\ket{2-222}, \ket{2-22-2},\\ \ket{2-2-22}, \ket{2-2-2-2}, \ket{-2222}, \ket{-222-2}, \ket{-22-22}, \ket{-22-2-2}, \ket{-2-222}, \ket{-2-22-2},\\ \ket{-2-2-22}, \ket{-2-2-2-2} \}$ with all possible measurements for the coin basis  as discussed in Table.~\ref{tab3}. For each set of bases defined from $P_{1}$ to $P_{15}$, a table, as shown in Table.~\ref{tab3}, can be easily constructed to complete the teleportation protocol. Because of the repetition of the above procedure, we keep it as an exercise for the readers. 
\section{Bidirectional teleportation of two qubit using two step quantum walk \label{sec5}}
\begin{figure}
    \begin{center}
    \includegraphics[scale=1.3]{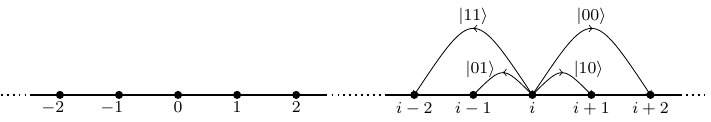}
    \end{center}
    \caption{The quantum walk on a one dimensional line using two coins. If  coins outcome is $00$, the walker moves from $i$ to $i+2$; if coins outcomes is $11$, the walker moves from $i$ to $i-2$; if coins outcomes is $10$, the walker moves from $i$ to $i+1$; if coin outcomes is $01$, the walker moves from $i$ to $i-1$.}
    \label{Figure5}
    \end{figure}

Bidirectional two qubit teleportation, as discussed in the last section, can also be performed using single and two step quantum walks. In this case, single and two step quantum walks are defined in terms of the combined outcomes of two coins as follows as shown in FIG~\ref{Figure5}: If the outcomes of the two coins are
$00$, the Walker moves two steps forward; if the outcome is $11$, the Walker moves two steps backward; if the outcome
is $01$, the Walker moves one step forward; and if the outcome is $10$, the Walker moves one step backward. 

Here also, the unknown state that Alice wants to send to Bob is $|\phi_{1}\rangle=a_{0}\ket{00}+a_{1}\ket{01}+a_{2}\ket{10}+a_{3}\ket{11}$  and the unknown state that Bob wants to send to Alice is $|\phi_{2}\rangle=b_{0}\ket{00}+b_{1}\ket{01}+b_{2}\ket{10}+b_{3}\ket{11}$. Let Alice have five particles: $A_{1}, A_{2}$,...$A_{5}$, where the state of the particle $A_{1}$ represents the position spaces of Alice, $P_{A_{1}}$. The states of particles  $A_{2}, A_{3}, A_{4}$ and $A_{5}$ represent the coin spaces of the coins $C_{1}, C_{2}, C_{3}$ and $C_{4}$, respectively. Similarly, Bob also has five particles: $B_{1}, B_{2}$,...$B_{5}$, where the state of the   particle $B_{1}$ represents the position space of Alice  $P_{B_{1}}$. As with Alice, states of particles  $B_{2}, B_{3}, B_{4}$, and $B_{5}$ represent the coin spaces of the coins $C_{5}, C_{6}, C_{7}$ and, $C_{8}$, respectively.

Initially, the unknown two qubit state to be teleported by Alice is encoded in $A_{2}$ and $A_{3}$, and the unknown two qubit state to be teleported by Bob is encoded in the $B_{2}$ and $B_{3}$. Also,  initialize the position spaces of $A_{1}$ and $B_{1}$  to $\ket{0}$, and initialize the coin spaces of $A_{4}$, $A_{5}$ and  $B_{4}$, $B_{5}$ to $\ket{00}$. Then the total initial state, in the convenient order is
\begin{equation}
    \begin{split}
\ket{\psi_{0}}_{A_{1}B_{1}A_{2}A_{3}A_{4}A_{5}B_{2}B_{3}B_{4}B_{5}} = \ket{00}\otimes(a_{0}\ket{00}+a_{1}\ket{01}+a_{2}\ket{10}+a_{3}\ket{11})\otimes\ket{00}\otimes\\(b_{0}\ket{00}+b_{1}\ket{01}+b_{2}\ket{10}+b_{3}\ket{11})\otimes\ket{00}
\end{split}
\label{Eqs4-1}
\end{equation}
The conditional shift operators are the fundamental elements of any quantum walk teleportation protocol, and in this case which can be described as  
 \begin{equation}
    \begin{split}
S_{1} =\sum_{n}(\ket{n+1}\bra{n})\\
S_{1}^\dagger=\sum_{n}(\ket{n}\bra{n+1})\\
S_{2} =\sum_{n}(\ket{n+2}\bra{n})\\
S_{2}^\dagger=\sum_{n}(\ket{n}\bra{n+2}).
\end{split}
\label{Eqs4-2}
\end{equation}
The bidirectional quantum teleportation process involves four walks, which are
\begin{equation}
    \begin{split}
W_1 =&E_{1}(I_{A_{1}B_{1}}\otimes I_{A_{2}A_{3}}\otimes I_{A_{4}A_{5}}\otimes I_{B_{2}B_{3}}\otimes I_{B_{4}B_{5}})\\
W_{2}=&E_{2}(I_{A_{1}B_{1}}\otimes I_{A_{2}A_{3}}\otimes I_{A_{4}A_{5}}\otimes I_{B_{2}B_{3}}\otimes I_{B_{4}B_{5}})  \\
W_3=&E_3 (I_{A_{1}B_{1}}\otimes I_{A_{2}A_{3}}\otimes (H \otimes H)_{A_{4}A_{5}}\otimes I_{B_{2}B_{3}}\otimes I_{B_{4}B_{5}})\\
W_{4}=&E_{4} (I_{A_{1}B_{1}}\otimes I_{A_{2}A_{3}}\otimes I_{A_{4}A_{5}}\otimes I_{B_{2}B_{3}}\otimes (H \otimes H)_{B_{4}B_{5}})
\end{split}
\label{Eqs4-3}
\end{equation}
where,
\begin{equation}
    \begin{split}
E_{1}=S_{1}\otimes  I_{B_{1}}\otimes \ket{01}\bra{01} \otimes I_{A_{4}A_{5}} \otimes I_{B_{2}B_{3}} \otimes I_{B_{4}B_{5}} +S_{1}^\dagger \otimes I_{B_{1}}\otimes  \ket{10}\bra{10} \otimes I_{A_{4}A_{5}}\otimes I_{B_{2}B_{3}}\otimes I_{B_{4}B_{5}}+\\S_{2}\otimes I_{B_{1}}\otimes \ket{00}\bra{00} \otimes I_{A_{4}A_{5}} \otimes I_{B_{2}B_{3}} \otimes I_{B_{4}B_{5}}+S_{2}^\dagger \otimes I_{B_{1}}\otimes \ket{11}\bra{11} \otimes I_{A_{4}A_{5}}\otimes I_{B_{2}B_{3}}\otimes I_{B_{4}B_{5}},\\
E_{2} = I_{A_{1}} \otimes S_{1}\otimes I_{A_{2}A_{3}}\otimes I_{A_{4}A_{5}} \otimes \ket{01}\bra{01} \otimes I_{B_{4}B_{5}}+I_{A_{1}}\otimes S_{1}^\dagger \otimes I_{A_{2}A_{3}}\otimes I_{A_{4}A_{5}} \otimes \ket{10}\bra{10} \otimes I_{B_{4}B_{5}}+\\I_{A_{1}}\otimes S_{2}\otimes I_{A_{2}A_{3}}\otimes I_{A_{4}A_{5}} \otimes \ket{00}\bra{00} \otimes I_{B_{4}B_{5}}+ I_{A_{1}}\otimes S_{2}^\dagger \otimes I_{A_{2}A_{3}}\otimes I_{A_{4}A_{5}} \otimes \ket{11}\bra{11} \otimes I_{B_{4}B_{5}},\\
E_{3}=I_{A_1} \otimes S_{1}\otimes I_{A_{2}A_{3}} \otimes \ket{01}\bra{01} \otimes I_{B_{2}B_{3}}\otimes I_{B_{4}B_{5}} + I_{A_1}\otimes S_{1}^\dagger \otimes  I_{A_{2}A_{3}}\otimes \ket{10}\bra{10} \otimes I_{B_{2}B_{3}}\otimes I_{B_{4}B_{5}} + \\I_{A_1}\otimes S_{2}\otimes I_{A_{2}A_{3}} \otimes \ket{00}\bra{00} \otimes  I_{B_{2}B_{3}}\otimes I_{B_{4}B_{5}} + I_{A_1} \otimes S_{2}^\dagger \otimes I_{A_{2}A_{3}}\otimes \ket{11}\bra{11} \otimes I_{B_{2}B_{3}}\otimes I_{B_{4}B_{5}}, \\
E_{4} =S_{1}\otimes  I_{B_1}\otimes I_{A_{2}A_{3}}\otimes I_{A_{4}A_{5}}\otimes I_{B_{2}B_{3}}\otimes \ket{01}\bra{01} + S_{1}^\dagger \otimes I_{B_1}\otimes I_{A_{2}A_{3}}\otimes I_{A_{4}A_{5}}\otimes I_{B_{2}B_{3}} \otimes \ket{10}\bra{10}+\\ S_{2}\otimes  I_{B_1}\otimes I_{A_{2}A_{3}}\otimes I_{A_{4}A_{5}}\otimes I_{B_{2}B_{3}} \otimes  \ket{00}\bra{00}+ S_{2}^\dagger \otimes  I_{B_1}\otimes I_{A_{2}A_{3}}\otimes I_{A_{4}A_{5}}\otimes I_{B_{2}B_{3}} \otimes \ket{11}\bra{11}.
\end{split}
\label{Eqs4-4}
\end{equation}
In the above expression, the notation $I_{A_{1}A_{2}}$ is used in place of $I_{A_{1}} \otimes I_{A_{2}}$. After performing the four different quantum walks defined in Eqs. (\ref{Eqs4-2}), (\ref{Eqs4-3}), and (\ref{Eqs4-4}), on the initial state given by Eq.~(\ref{Eqs4-1}), we get an expression given in \text{Appendix}. (\ref{Eqn4final}).

\begin{figure}
    \begin{center}
    \includegraphics[scale=.65]{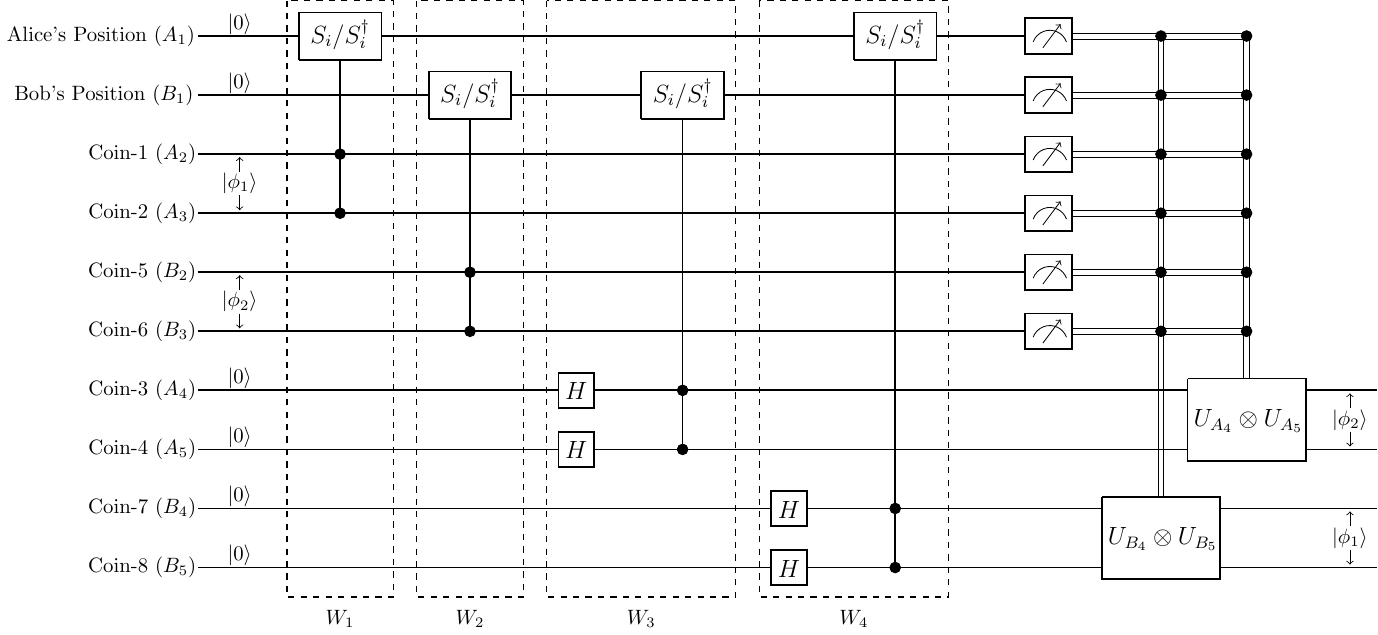}
    \end{center}
    \caption{Circuit diagram for bidirectional teleportation of two-qubit states using single and two steps quantum walk.}
    \label{Figure6}
    \end{figure}

If Alice and Bob jointly measure $|00\rangle_{A_{1}B_{1}}$ in their position spaces, the resulting state is
\begin{equation}
    \begin{split}
    \ket{\psi_{2}}=&\frac{1}{4}(a_{0}b_{0}\ket{00110011}+a_{0}b_{1}\ket{00100111}+a_{0}b_{2}\ket{00011011}+a_{0}b_{3}\ket{00001111}+\\ & a_{1}b_{0}\ket{01110010}+a_{1}b_{1}\ket{01100110}+a_{1}b_{2}\ket{01011010}+a_{1}b_{3}\ket{01001110}+\\ & a_{2}b_{0}\ket{10110001}+a_{2}b_{1}\ket{10100101}+a_{2}b_{2}\ket{10011001}+a_{2}b_{3}\ket{10001101}+\\ & a_{3}b_{0}\ket{11110000}+a_{3}b_{1}\ket{11100100}+a_{3}b_{2}\ket{11011000}+a_{3}b_{3}\ket{11001100})
        \end{split}
\end{equation}
and Alice performs a measurement in her coin bases  $\ket{++}_{A_{2}A_{3}}$ and Bob performs a measurement in his coin bases $\ket{++}_{B_{2}B_{3}}$, then the final state will be
\begin{equation}
    \begin{split}
    \ket{\psi_{3}}=&\frac{1}{16}(a_{0}b_{0}\ket{1111}+a_{0}b_{1}\ket{1011}+a_{0}b_{2}\ket{0111}+a_{0}b_{3}\ket{0011}+a_{1}b_{0}\ket{1110}+a_{1}b_{1}\ket{1010}+ \\ & a_{1}b_{2}\ket{0110}+a_{1}b_{3}\ket{0010}+a_{2}b_{0}\ket{1101}+a_{2}b_{1}\ket{1001}+a_{2}b_{2}\ket{0101}+a_{2}b_{3}\ket{0001}\\& +a_{3}b_{0}\ket{1100}+a_{3}b_{1}\ket{1000}+a_{3}b_{2}\ket{0100}+a_{3}b_{3}\ket{0000})
        \end{split}
\end{equation}
which can be rewritten as
\begin{equation}
    \begin{split}
    \ket{\psi_{4}}_{A_{4}A_{5}B_{4}B_{5}}=&\frac{1}{16}\big((b_{0}\ket{11}+b_{1}\ket{10}+b_{2}\ket{01}+b_{3}\ket{00})_{A_{4}A_{5}}\otimes(a_{0}\ket{11}+a_{1}\ket{10}+ \\ & a_{2}\ket{01}+a_{3}\ket{00})_{B_{4}B_{5}}\big)\\
    \end{split}
\end{equation}
Then, applying the $(\sigma_{x})_{A_{4}}(\sigma_{x})_{A_{5}}(\sigma_{x})_{B_{4}}(\sigma_{x})_{B_{5}}$ operations on the above state results in
\begin{equation}
    \begin{split}
    \ket{\psi_{5}}_{A_{4}A_{5}B_{4}B_{5}}=&\frac{1}{16}\big((b_{0}\ket{00}+b_{1}\ket{01}+b_{2}\ket{10}+b_{3}\ket{11})_{A_{4}A_{5}}\otimes(a_{0}\ket{00}+a_{1}\ket{01}+ \\ & a_{2}\ket{10}+a_{3}\ket{11})_{B_{4}B_{5}}\big)\\
    \end{split}
\end{equation}
That is Alice is able to teleport unknown state to Bob and Bob is able to teleport unknown state to Alice simultaneously. Table.\ref{tab4} shows the remaining measurements on both Alice's and Bob's coins and the corresponding unitary operations for bidirectional teleportation. FIG.~\ref{Figure6} describes the two-qubit bidirectional teleportation protocol using single- and two-step quantum walks. By comparing Table.\ref{tab3} and Table.\ref{tab4}, we observe that both give the same results when the position measurement is $|0000\rangle$ in the single-step quantum walk and $|00\rangle$ in the two-step quantum walk.

\begin{table}
\caption{Outcomes of the measurements performed by Alice and Bob in their position basis $|00\rangle$, and the first two coin bases of both, along with the corresponding unitary operations on their last two coins, for reproducing the unknown two-qubit states.}
\begin{tabular}{c  c c}
\hline
position basis~~~~~~~~  & ~~~~ coin basis  ~~~~&~ Unitary operation ~~ \\[1.ex]
\hline
 & $\ket{++}_{A_{2}A_{3}}\ket{++}_{B_{2}B_{3}}$ & $\sigma_{A_{4}}^{x}\sigma_{A_{5}}^{x}\sigma_{B_{4}}^{x}\sigma_{B_{5}}^{x}$  \\
& $\ket{++}_{A_{2}A_{3}}\ket{+-}_{B_{2}B_{3}}$ & $\sigma_{A_{5}}^{z}\sigma_{A_{4}}^{x}\sigma_{A_{5}}^{x}\sigma_{B_{4}}^{x}\sigma_{B_{5}}^{x}$ \\
& $\ket{++}_{A_{2}A_{3}}\ket{-+}_{B_{2}B_{3}}$ & $\sigma_{A_{4}}^{z}\sigma_{A_{4}}^{x}\sigma_{A_{5}}^{x}\sigma_{B_{4}}^{x}\sigma_{B_{5}}^{x}$ \\
& $\ket{++}_{A_{2}A_{3}}\ket{--}_{B_{2}B_{3}}$ & $\sigma_{A_{5}}^{z}\sigma_{A_{4}}^{z}\sigma_{A_{4}}^{x}\sigma_{A_{5}}^{x}\sigma_{B_{4}}^{x}\sigma_{B_{5}}^{x}$ \\
& $\ket{+-}_{A_{2}A_{3}}$$\ket{++}_{B_{2}B_{3}}$ &$\sigma_{A_{4}}^{x}\sigma_{A_{5}}^{x}\sigma_{B_{5}}^{z}\sigma_{B_{4}}^{x}\sigma_{B_{5}}^{x}$ \\
& $\ket{+-}_{A_{2}A_{3}}\ket{+-}_{B_{2}B_{3}}$ & $\sigma_{A_{5}}^{z}\sigma_{A_{4}}^{x}\sigma_{A_{5}}^{x}\sigma_{B_{5}}^{z}\sigma_{B_{4}}^{x}\sigma_{B_{5}}^{x}$ \\
& $\ket{+-}_{A_{2}A_{3}}\ket{-+}_{B_{2}B_{3}}$ & $\sigma_{A_{4}}^{z}\sigma_{A_{4}}^{x}\sigma_{A_{5}}^{x}\sigma_{B_{5}}^{z}\sigma_{B_{4}}^{x}\sigma_{B_{5}}^{x}$ \\
$\ket{00}_{A_{1}B_{1}}$ & $\ket{+-}_{A_{2}A_{3}}\ket{--}_{B_{2}B_{3}}$ & $\sigma_{A_{5}}^{z}\sigma_{A_{4}}^{z}\sigma_{A_{4}}^{x}\sigma_{A_{5}}^{x}\sigma_{B_{5}}^{z}\sigma_{B_{4}}^{x}\sigma_{B_{5}}^{x}$ \\
& $\ket{-+}_{A_{2}A_{3}}\ket{++}_{B_{2}B_{3}}$ & $\sigma_{A_{4}}^{x}\sigma_{A_{5}}^{x}\sigma_{B_{4}}^{z}\sigma_{B_{4}}^{x}\sigma_{B_{5}}^{x}$\\
& $\ket{-+}_{A_{2}A_{3}}\ket{+-}_{B_{2}B_{3}}$ & $\sigma_{A_{5}}^{z}\sigma_{A_{4}}^{x}\sigma_{A_{5}}^{x}\sigma_{B_{4}}^{z}\sigma_{B_{4}}^{x}\sigma_{B_{5}}^{x}$ \\
& $\ket{-+}_{A_{2}A_{3}}\ket{-+}_{B_{2}B_{3}}$ & $\sigma_{A_{4}}^{z}\sigma_{A_{4}}^{x}\sigma_{A_{5}}^{x}\sigma_{B_{4}}^{z}\sigma_{B_{4}}^{x}\sigma_{B_{5}}^{x}$\\
& $\ket{-+}_{A_{2}A_{3}}\ket{--}_{B_{2}B_{3}}$ & $\sigma_{A_{5}}^{z}\sigma_{A_{4}}^{z}\sigma_{A_{4}}^{x}\sigma_{A_{5}}^{x}\sigma_{B_{4}}^{z}\sigma_{B_{4}}^{x}\sigma_{B_{5}}^{x}$\\
& $\ket{--}_{A_{2}A_{3}}\ket{++}_{B_{2}B_{3}}$ & $\sigma_{A_{4}}^{x}\sigma_{A_{5}}^{x}\sigma_{B_{5}}^{z}\sigma_{B_{4}}^{z}\sigma_{B_{4}}^{x}\sigma_{B_{5}}^{x} $ \\
& $\ket{--}_{A_{2}A_{3}}\ket{-+}_{B_{2}B_{3}}$ & $\sigma_{A_{4}}^{z}\sigma_{A_{4}}^{x}\sigma_{A_{5}}^{x}\sigma_{B_{5}}^{z}\sigma_{B_{4}}^{z}\sigma_{B_{4}}^{x}\sigma_{B_{5}}^{x}$\\
& $\ket{--}_{A_{2}A_{3}}\ket{--}_{B_{2}B_{3}}$ & $\sigma_{A_{5}}^{z}\sigma_{A_{4}}^{z}\sigma_{A_{4}}^{x}\sigma_{A_{5}}^{x}\sigma_{B_{5}}^{z}\sigma_{B_{4}}^{z}\sigma_{A_{4}}^{x}\sigma_{B_{5}}^{x}$ \\
[1.ex]
\hline
\end{tabular}
\label{tab4}
\end{table}

Similarly, measurements can be performed using the orthonormal states formed from the following disjoint set of states: $Q_{1}=\{ \ket{10}, \ket{-10} \},~ Q_{2}=\{\ket{30}, \ket{-30} \},~Q_{3}=\{\ket{40}, \ket{20}, \ket{-20}, \ket{-40} \},~\\ Q_{4}=\{\ket{01}, \ket{0-1}\},~ Q_{5}=\{\ket{11}, \ket{1-1}, \ket{-11}, \ket{-1-1}\},~ Q_{6}=\{\ket{31}, \ket{3-1}, \ket{-31}, \ket{-3-1} \},~\\ Q_{7}=\{\ket{41}, \ket{4-1}, \ket{21}, \ket{2-1}, \ket{-21}, \ket{-2-1}, \ket{-41}, \ket{-4-1} \},~ Q_{8}=\{\ket{03},\ket{0-3} \},~ Q_{9}=\ket{13}, \ket{1-3}, \ket{-13}, \ket{-1-3} \},~ Q_{10}=\{\ket{33}, \ket{3-3}, \ket{-33}, \ket{-3-3} \},~ Q_{11}=\{\ket{43}, \ket{4-3}, \ket{23},\\ \ket{2-3}, \ket{-23}, \ket{-2-3}, \ket{-43}, \ket{-4-3} \},~ Q_{12}=\{\ket{04}, \ket{02}, \ket{0-2}, \ket{0-4} \},~ Q_{13}=\{\ket{14}, \ket{12},\\ \ket{1-2}, \ket{1-4}, \ket{-14}, \ket{-12}, \ket{-1-2}, \ket{-1-4} \},~Q_{14}=\{\ket{34}, \ket{32}, \ket{3-2}, \ket{3-4}, \ket{-34},\\ \ket{-32}, \ket{-3-2}, \ket{-3-4} \},~ Q_{15}= \{ \ket{44}, \ket{42}, \ket{4-2}, \ket{4-4}, \ket{24}, \ket{22}, \ket{2-2}, \ket{2-4}, \ket{-24},\\ \ket{-22}, \ket{-2-2}, \ket{-2-4}, \ket{-44}, \ket{-42}, \ket{-4-2}, \ket{-4-4} \}$.

We can easily construct tables for each set of bases from $Q_{1}$ to $Q_{15}$ as shown in Table.~\ref{tab4}.  A comparison of the equivalent tables generated for the set of bases from $P_{1}$ to $P_{15}$ indicates that the equivalent sets of bases yield the same result, as evident from the comparison between Table.~\ref{tab3} and Table.~\ref{tab4}. It is conclusively determined from these results that two-qubit teleportation via single-step and two-step quantum walks is equivalent.

\section{Conclusion \label{sec6}}
We studied single-qubit and two-qubit bidirectional teleportation using quantum walk techniques. In this teleportation scheme, initial entanglement is not required and can be generated in the four steps of the quantum walk process. In bidirectional single-qubit teleportation, nearest-neighbor jumps are introduced based on the outcome of a single coin. By implementing such jumps on two independent one-dimensional lattices, a protocol for single-qubit bidirectional teleportation is developed. A bidirectional single-qubit teleportation scheme is also developed using quantum walks on a four-vertex circle ($4$-cycle). All possible measurement outcomes and the corresponding unitary transformations required to reconstruct the teleported state in both cases have been tabulated.

We also studied bidirectional two-qubit teleportation using two methods: the first involves nearest-neighbor jumps on a one-dimensional lattice based on the outcome of a coin, and the second involves both nearest-neighbor and next-nearest-neighbor jumps based on the outcomes of two coins. Two distinct protocols for two-qubit bidirectional teleportation are developed. A comparison between the two protocols reveals a one-to-one mapping, demonstrating their equivalence. 

These teleportation schemes can be extended to a larger number of qubits using single-step quantum walks and combinations of multi-step quantum walks. However, our teleportation schemes have certain limitations, particularly in remote scenarios, as the quantum walk steps require joint operations by the sender and receiver—an arrangement that becomes challenging when they are spatially separated. Addressing these challenges remains an important open problem in this area of research.
\appendix

\section{Final state after a four-step quantum walk in bidirectional teleportation of two qubits using a single-step quantum walk. \label{AppA}}
In this teleportation scheme, after the four walks defined in Eqs. (\ref{Eqn5-2}) and (\ref{Eqn5-3}) act on the input state in Eq. (\ref{Eqn5-1}), it becomes

\begin{align}
    |\psi_1\rangle=& \frac{1}{4} \Big(a_{0}b_{0}|222200000000\rangle+a_{0}b_{0}|202200000001\rangle+a_{0}b_{0}|022200000010\rangle+a_{0}b_{0}|002200000011\rangle + \nonumber\\ &a_{0}b_{0}|222000010000\rangle+a_{0}b_{0}|202000010001\rangle+a_{0}b_{0}|022000010010\rangle+a_{0}b_{0}|002000010011\rangle +\nonumber\\&
a_{0}b_{0}|220200100000\rangle+a_{0}b_{0}|200200100001\rangle+a_{0}b_{0}|020200100010\rangle+a_{0}b_{0}|000200100011\rangle + \nonumber\\
&a_{0}b_{0}|220000110000\rangle+a_{0}b_{0}|200000110001\rangle+a_{0}b_{0}|020000110010\rangle+a_{0}b_{0}|000000110011\rangle+\nonumber\\
& a_{0}b_{1}|222000000100\rangle + a_{0}b_{1}|202000000101\rangle +a_{0}b_{1}|022000000110\rangle+\nonumber\\
&a_{0}b_{1}|002000000111\rangle +a_{0}b_{1}|222-200010100\rangle+a_{0}b_{1}|202-200010101\rangle+\nonumber\\
&a_{0}b_{1}|022-200010110\rangle+a_{0}b_{1}|002-200010111\rangle+a_{0}b_{1}|220000100100\rangle+\nonumber\\
&a_{0}b_{1}|200000100101\rangle+a_{0}b_{1}|020000100110\rangle+a_{0}b_{1}|000000100111\rangle+ a_{0}b_{1}|220-200110100\rangle+\nonumber\\
&a_{0}b_{1}|200-200110101\rangle+a_{0}b_{1}|020-200110110\rangle+a_{0}b_{1}|000-200110111\rangle+\nonumber\\
&a_{0}b_{2}|220200001000\rangle+a_{0}b_{2}|200200001001\rangle+a_{0}b_{2}|020200001010\rangle+a_{0}b_{2}|000200001011\rangle + \nonumber\\
& a_{0}b_{2}|220000011000\rangle+a_{0}b_{2}|200000011001\rangle+a_{0}b_{2}|020000011010\rangle+a_{0}b_{2}|000000011011\rangle+\nonumber\\
&
a_{0}b_{2}|22-2200101000\rangle+a_{0}b_{2}|20-2200101001\rangle+a_{0}b_{2}|02-2200101010\rangle+\nonumber\\
&a_{0}b_{2}|00-2200101011\rangle+a_{0}b_{2}|22-2000111000\rangle+a_{0}b_{2}|20-2000111001\rangle+\nonumber\\
&a_{0}b_{2}|02-2000111010\rangle+a_{0}b_{2}|00-2000111011\rangle+a_{0}b_{3}|220000001100\rangle + \nonumber\\
& a_{0}b_{3}|200000001101\rangle+ a_{0}b_{3}|020000001110\rangle+a_{0}b_{3}|000000001111\rangle+  a_{0}b_{3}|220-200 011100\rangle \nonumber\\
& +a_{0}b_{3}|200-200011101\rangle+a_{0}b_{3}|020-200011110\rangle+  a_{0}b_{3}|000-200011111\rangle+\nonumber\\
& a_{0}b_{3}|22-2000101100\rangle +a_{0}b_{3}|20-2000 101101\rangle+a_{0}b_{3}|02-2000101110\rangle+\nonumber\\
& a_{0}b_{3}|00-2000101111\rangle+a_{0}b_{3}|22-2-200111100\rangle+a_{0}b_{3}|20-2-200111101\rangle+\nonumber\\
& a_{0}b_{3}|02-2-200111110\rangle+a_{0}b_{3}|00-2-200111111\rangle+a_{0}b_{3}|22-2101100\rangle+\nonumber\\
&a_{0}b_{3}|20-2000101101\rangle+a_{0}b_{3}|02-2000101110\rangle+a_{0}b_{3}|00-2000101111\rangle+\nonumber\\
&a_{0}b_{3}|22-2-200111100\rangle+a_{0}b_{3}|20-2-200111101\rangle+a_{0}b_{3}|02-2-200111110\rangle+\nonumber\\&a_{0}b_{3}|00-2-200111111\rangle+a_{1}b_{0}|202201000000\rangle+a_{1}b_{0}|2-22201000001\rangle+ \nonumber\\&a_{1}b_{0}|002201000010\rangle+a_{1}b_{0}|0-22201000011\rangle+a_{1}b_{0}|202001010000\rangle + \nonumber \\ & a_{1}b_{0}|2-22001010001\rangle+a_{1}b_{0}|002001010010\rangle+a_{1}b_{0}|0-22001010011\rangle+\nonumber\\&a_{1}b_{0}|200201100000\rangle+a_{1}b_{0}|2-20201100001\rangle+a_{1}b_{0}|000201100010\rangle+a_{1}b_{0}|0-20201100011\rangle+\nonumber\\& a_{1}b_{0}|200001110000\rangle+a_{1}b_{0}|2-20001110001\rangle+a_{1}b_{0}|000001110010\rangle+a_{1}b_{0}|0-20001110011\rangle+\nonumber\\&a_{1}b_{1}|202001000100\rangle+a_{1}b_{1}|2-22001000101\rangle+a_{1}b_{1}|002001000110\rangle+a_{1}b_{1}|0-22001000111\rangle+\nonumber\\&a_{1}b_{1}|202-201010110\rangle+a_{1}b_{1}|2-22-201010101\rangle+a_{1}b_{1}|002-201010110\rangle+\nonumber\\&
a_{1}b_{1}|0-22-201010111\rangle+a_{1}b_{1}|200001100100\rangle+a_{1}b_{1}|2-20001100101\rangle+\nonumber \\ &a_{1}b_{1}|000001100110\rangle+a_{1}b_{1}|0-20001100111\rangle+a_{1}b_{1}|200-21110100\rangle+\nonumber\\&a_{1}b_{1}|2-20-201110101\rangle+a_{1}b_{1}|00-201110110\rangle+a_{1}b_{1}|0-20-201110111\rangle+\nonumber\\&a_{1}b_{2}|200201001000\rangle+a_{1}b_{2}|220201001001\rangle+a_{1}b_{2}|000201001010\rangle+a_{1}b_{2}|0-20201001001\rangle+\nonumber\\&a_{1}b_{2}|200001011000\rangle+a_{1}b_{2}|2-20001011001\rangle+a_{1}b_{2}|000001011010\rangle+a_{1}b_{2}|0-20010111011\rangle \nonumber\\&+a_{1}b_{2}|20-2201101000\rangle+a_{1}b_{2}|2-2-220110101\rangle+a_{1}b_{2}|00-2201101010\rangle+\nonumber\\&a_{1}b_{2}|0-2-2201101011\rangle+a_{1}b_{2}|20-2001111000\rangle+a_{1}b_{2}|2-2-200111001\rangle+\nonumber\\&a_{1}b_{2}|00-2001111010\rangle+a_{1}b_{2}|0-2-2001111011\rangle+a_{1}b_{3}|200001001100\rangle+\nonumber \\ & a_{1}b_{3}
        |2-20001001101\rangle+a_{1}b_{3}|000001001110\rangle+a_{1}b_{3}|0-20001001111\rangle+\nonumber \\ & a_{1}b_{3}|200-201011100\rangle+a_{1}b_{3}|2-20-201011101\rangle000-201011110\rangle|0-20-201011111\rangle+\nonumber\\&a_{1}b_{3}|20-2001101100\rangle+a_{1}b_{3}|2-2-2001101101\rangle+a_{1}b_{3}|00-2001101100\rangle+\nonumber\\&a_{1}b_{3}|0-2-2-201111111\rangle+a_{1}b_{3}|20-2-201111100\rangle+a_{1}b_{3}|2-2-2-201111101\rangle+\nonumber\\& a_{1}b_{3}|00-2-201111110\rangle+a_{1}b_{3}|0-2-2-201111111\rangle+a_{2}b_{0}|022210000000\rangle+\nonumber\\&a_{2}b_{0}|002210000001\rangle+a_{2}b_{0}|-222210000010\rangle+a_{2}b_{0}|-202210000011\rangle+\nonumber\\&a_{2}b_{0}|022010010000\rangle+a_{2}b_{0}|002010010001\rangle+a_{2}b_{0}|-222010010010\rangle+a_{2}b_{0}|-202010010011\rangle+\nonumber\\&a_{2}b_{0}|020210100000\rangle+a_{2}b_{0}|000210100001\rangle+a_{2}b_{0}|-220210100010\rangle+a_{2}b_{0}|-20021010011\rangle+\nonumber\\&a_{2}b_{0}|020010110000\rangle+a_{2}b_{0}|000010110001\rangle+a_{2}b_{0}|-220010110010\rangle+a_{2}b_{0}|-200010110011\rangle+\nonumber\\
        &a_{2}b_{1}|022010000100\rangle+a_{2}b_{1}|002010000101\rangle+a_{2}b_{1}|-222010000110\rangle+a_{2}b_{1}|-202010000111\rangle+\nonumber\\&a_{2}b_{1}|022-210010100\rangle+a_{2}b_{1}|002-210010101\rangle+a_{2}b_{1}|-222-21001011\rangle+\rangle+\nonumber\\&a_{2}b_{1}|-202-210010111\rangle+a_{2}b_{1}|020010100100\rangle+a_{2}b_{1}|000010100101\rangle+\nonumber\\&a_{2}b_{1}|-220010100110\rangle+a_{2}b_{1}|-200010100111\rangle+a_{2}b_{1}|020-210110100\rangle+\nonumber\\& a_{2}b_{1}|000-210110101\rangle+a_{2}b_{1}|-220-210110110\rangle+a_{2}b_{1}|-200-210110111\rangle+\nonumber\\&a_{2}b_{2}|020210001000\rangle+a_{2}b_{2}|000210001001\rangle+a_{2}b_{2}|-220210001010\rangle+a_{2}b_{2}|-200210001011\rangle \nonumber\\&+a_{2}b_{2}|020010011000\rangle+a_{2}b_{2}|000010011001\rangle+a_{2}b_{2}|-220010011010\rangle+a_{2}b_{2}|-200010011011\rangle \nonumber\\&+a_{2}b_{2}|02-2210101000\rangle+a_{2}b_{2}|00-2210101001\rangle+a_{2}b_{2}|-22-2210101010\rangle+\nonumber\\&a_{2}b_{2}|-20-2210101011\rangle+a_{2}b_{2}|02-2010111000\rangle+a_{2}b_{2}|00-2010111001\rangle+\nonumber\\&a_{2}b_{2}|-22-2010111010\rangle+a_{2}b_{2}|-20-2010111011\rangle+a_{2}b_{3}|020010001100\rangle +\nonumber \\&a_{2}b_{3}|000010001101\rangle+ a_{2}b_{3}|-220010001110\rangle + a_{2}b_{3}|-200010001111\rangle+\nonumber\\&a_{2}b_{3}|020-210011100\rangle+ a_{2}b_{3}|000-210011101\rangle+ a_{2}b_{3}|-220-210011110\rangle + \nonumber \\& a_{2}b_{3}|-200-210011111\rangle+a_{2}b_{3}|02-2010101100\rangle+ a_{2}b_{3}|00-2010101101\rangle+\nonumber \\& a_{2}b_{3}|-22-2010101110\rangle+ a_{2}b_{3}|-20-2010101111\rangle+ a_{2}b_{3}|22-2-210111100\rangle +\nonumber \\& a_{2}b_{3}|00-2-210111101\rangle + a_{2}b_{3}|-22-2-210111110\rangle + a_{2}b_{3}|-20-2-220111111\rangle+\nonumber\\&a_{3}b_{0}|002211000000\rangle+a_{3}b_{0}|0-22211000001\rangle+a_{3}b_{0}|-202211000010\rangle+\nonumber \\ & a_{3}b_{0}|-2-22211000011\rangle+a_{3}b_{0}|002211010000\rangle+a_{3}b_{0}|0-22011010001\rangle+\nonumber\\&a_{3}b_{0}|-202011010010\rangle+a_{3}b_{0}|-2-22011010011\rangle+a_{3}b_{0}|000211100000\rangle+\nonumber\\&a_{3}b_{0}|0-20211100001\rangle+a_{3}b_{0}|-200211100010\rangle+a_{3}b_{0}|-2-20211100011\rangle+\nonumber\\&a_{3}b_{0}|000011110000\rangle+a_{3}b_{0}|0-20011110001\rangle+a_{3}b_{0}|-200011110010\rangle+\nonumber\\&a_{3}b_{0}|-2-20011110011\rangle+a_{3}b_{1}|002011000100\rangle+a_{3}b_{1}|0-22011000101\rangle+\nonumber\\&a_{3}b_{1}|-202011000110\rangle+a_{3}b_{1}|-2-22011000111\rangle+a_{3}b_{1}|002-211010100\rangle+\nonumber\\&a_{3}b_{1}|0-22-211010101\rangle+a_{3}b_{1}|-202-211010110\rangle+a_{3}b_{1}|-2-22-211010111\rangle+\nonumber\\&a_{3}b_{1}|000011100111\rangle+a_{3}b_{1}|0-20011100101\rangle+a_{3}b_{1}|-200011100110\rangle+\nonumber\\&a_{3}b_{1}|-2-20011100111\rangle+a_{3}b_{1}|000-211110100\rangle+a_{3}b_{1}|0-20-211110101\rangle+\nonumber\\&a_{3}b_{1}|-200-211110110\rangle+a_{3}b_{1}|-2-20-211110111\rangle+a_{3}b_{2}|0002110010000\rangle+\nonumber\\&a_{3}b_{2}|0-20211001001\rangle+a_{3}b_{2}|-200211001010\rangle+a_{3}b_{2}|-2-20211001011\rangle+\nonumber\\&a_{3}b_{2}|000011011000\rangle+a_{3}b_{2}|0-20011011001\rangle+a_{3}b_{2}|-200011011010\rangle+\nonumber\\&a_{3}b_{2}|-2-20011011011\rangle+a_{3}b_{2}|00-2211101000\rangle+a_{3}b_{2}|0-2-2211101001\rangle+\nonumber\\&a_{3}b_{2}|-20-2211101010\rangle+a_{3}b_{2}|-2-2-2211101011\rangle+a_{3}b_{2}|00-2011111000\rangle+\nonumber\\&a_{3}b_{2}|0-2-2011111001\rangle+a_{3}b_{2}|-20-2011111010\rangle+a_{3}b_{2}|-2-2-211111011\rangle+\nonumber\\&a_{3}b_{3}|000011001100\rangle+a_{3}b_{3}|0-20011001101\rangle+a_{3}b_{3}|-200011001110\rangle+\nonumber\\&a_{3}b_{3}|-2-20011001111\rangle+a_{3}b_{3}|000-211011100\rangle+a_{3}b_{3}|0-20-211011101\rangle+\nonumber\\&a_{3}b_{3}|-200-211011110\rangle+a_{3}b_{3}|-2-20-211011111\rangle+a_{3}b_{3}|00-2011101100\rangle+\nonumber\\&a_{3}b_{3}|0-2-2011101101\rangle+a_{3}b_{3}|-20-2011101110\rangle+a_{3}b_{3}|-2-2-2011101111\rangle+\nonumber\\&a_{3}b_{3}|00-2211111100\rangle+a_{3}b_{3}|0-2-2-211111101\rangle+a_{3}b_{3}|-20-2-211111110\rangle+\nonumber\\&a_{3}b_{3}|-2-2-2-211111111\rangle \Big).
         \label{Eqn5final}
\end{align}
On the above state, measurements can be performed in the respective bases, and unitary operators can reproduce the two-qubit states in bidirectional teleportation.

\section{Final state after a four-step quantum walk in bidirectional teleportation of two qubit using two step quantum walk. \label{AppB}} 
As in the \textbf{Appendix A}, after four step walk defined in Eqs. (\ref{Eqs4-2}), (\ref{Eqs4-3}) and (\ref{Eqs4-4}) act on the input state in (\ref{Eqs4-1}) , it becomes
\begin{align}
&\ket{\psi_{1}}=\frac{1}{4} (a_{0}b_{0}\ket{4400000000}+a_{0}b_{0}\ket{3400000001}+a_{0}b_{0}\ket{1400000010}+a_{0}b_{0}\ket{0400000011}+ \nonumber \\ & a_{0}b_{0}\ket{4300010000}+a_{0}b_{0}\ket{3300010001}+a_{0}b_{0}\ket{1300010010}+a_{0}b_{0}\ket{0300010011}+ \nonumber  \\ & a_{0}b_{0}\ket{4100100000}+a_{0}b_{0}\ket{3100100001}+a_{0}b_{0}\ket{1100100010}+a_{0}b_{0}\ket{0100100011}+\nonumber   \\ & a_{0}b_{0}\ket{4000110000}+a_{0}b_{0}\ket{3000110001}+a_{0}b_{0}\ket{1000110010}+a_{0}b_{0}\ket{0000110011}+ \nonumber \\ &
a_{0}b_{1}\ket{4300000100}+a_{0}b_{1}\ket{3300000101}+a_{0}b_{1}\ket{1300000110}+a_{0}b_{1}\ket{0300000111}+\nonumber   \\ & a_{0}b_{1}\ket{4200010100}+a_{0}b_{1}\ket{3200010101}+a_{0}b_{1}\ket{1200010110}+a_{0}b_{1}\ket{0200010111}+\nonumber  \\ & a_{0}b_{1}\ket{4000100100}+a_{0}b_{1}\ket{3000100101}+a_{0}b_{1}\ket{1000100110}+a_{0}b_{1}\ket{0000100111}+ \nonumber \\ & a_{0}b_{1}\ket{4-100110100}+a_{0}b_{1}\ket{3-100110101}+a_{0}b_{1}\ket{1-100110110}+a_{0}b_{1}\ket{0-100110111}+ \nonumber \\&
a_{0}b_{2}\ket{4100001000}+a_{0}b_{2}\ket{3100001001}+a_{0}b_{2}\ket{1100001010}+a_{0}b_{2}\ket{0100001011}+\nonumber  \\ & a_{0}b_{2}\ket{4000011000}+a_{0}b_{2}\ket{3000011001}+a_{0}b_{2}\ket{1000011010}+a_{0}b_{2}\ket{0000011011}+ \nonumber \\ & a_{0}b_{2}\ket{4-200101000}+a_{0}b_{2}\ket{3-200101001}+a_{0}b_{2}\ket{1-200101010}+a_{0}b_{2}\ket{0-200101011}+\nonumber  \\ & 
a_{0}b_{2}\ket{4-300111000}+a_{0}b_{2}\ket{3-300111001}+a_{0}b_{2}\ket{1-300111010}+a_{0}b_{2}\ket{0-300111011}+\nonumber  \\ &
a_{0}b_{3}\ket{4000001100}+a_{0}b_{3}\ket{3000001101}+a_{0}b_{3}\ket{1000001110}+a_{0}b_{3}\ket{0000001111}+\nonumber  \\ & a_{0}b_{3}\ket{4 -100011100}+a_{0}b_{3}\ket{3-100011101}+a_{0}b_{3}\ket{1-100011110}+a_{0}b_{3}\ket{0-100011111}+\nonumber  \\ & 
a_{0}b_{3}\ket{4-300101100}+a_{0}b_{3}\ket{3-300101101}+a_{0}b_{3}\ket{1-300101110}+a_{0}b_{3}\ket{0-300101111}+\nonumber  \\ & 
a_{0}b_{3}\ket{4-400111100}+a_{0}b_{3}\ket{3-400111101}+a_{0}b_{3}\ket{1-400111110}+a_{0}b_{3}\ket{0-400111111}+ \nonumber \\ &
a_{1}b_{0}\ket{3401000000}+a_{1}b_{0}\ket{2401000001}+a_{1}b_{0}\ket{0401000010}+a_{1}b_{0}\ket{-1401000011}+ \nonumber \\ & a_{1}b_{0}\ket{3301010000}+a_{1}b_{0}\ket{2301010001}+a_{1}b_{0}\ket{0301010010}+a_{1}b_{0}\ket{-1301010011}+ \nonumber \\ & a_{1}b_{0}\ket{3101100000}+a_{1}b_{0}\ket{2101100001}+a_{1}b_{0}\ket{0101100010}+a_{1}b_{0}\ket{-1101100011}+ \nonumber \\ & a_{1}b_{0}\ket{3001110000}+a_{1}b_{0}\ket{2001110001}+a_{1}b_{0}\ket{0001110010}+a_{1}b_{0}\ket{-1001110011}+ \nonumber \\ & 
a_{1}b_{1}\ket{3301000100}+a_{1}b_{1}\ket{2301000101}+a_{1}b_{1}\ket{0301000110}+a_{1}b_{1}\ket{-1301000111}+ \nonumber \\ & a_{1}b_{1}\ket{3201010100}+a_{1}b_{1}\ket{2201010101}+a_{1}b_{1}\ket{0201010110}+a_{1}b_{1}\ket{-1201010111}+ \nonumber \\ & a_{1}b_{1}\ket{3001100100}+a_{1}b_{1}\ket{2001100101}+a_{1}b_{1}\ket{0001100110}+a_{1}b_{1}\ket{-1001100111}+ \nonumber \\ & a_{1}b_{1}\ket{3-101110100}+a_{1}b_{1}\ket{2-101110101}+a_{1}b_{1}\ket{0-101110110}+a_{1}b_{1}\ket{-1-101110111}+ \nonumber \\ & 
a_{1}b_{2}\ket{3101001000}+a_{1}b_{2}\ket{2101001001}+a_{1}b_{}\ket{0101001010}+a_{1}b_{2}\ket{-1101001011}+ \nonumber \\ & a_{1}b_{2}\ket{3001011000}+a_{1}b_{2}\ket{2001011001}+a_{1}b_{2}\ket{0001011010}+a_{1}b_{2}\ket{-1001011011}+ \nonumber \\ & a_{1}b_{2}\ket{3-201101000}+a_{1}b_{2}\ket{2-201101001}+a_{1}b_{2}\ket{0-201101010}+a_{1}b_{2}\ket{-1-201101011}+ \nonumber \\ & 
a_{1}b_{2}\ket{3-301111000}+a_{1}b_{2}\ket{2-301111001}+a_{1}b_{2}\ket{0-301111010}+a_{1}b_{2}\ket{-1-301111011}+ \nonumber \\ &
a_{1}b_{3}\ket{3001001100}+a_{1}b_{3}\ket{2001001101}+a_{1}b_{3}\ket{0001001110}+a_{1}b_{3}\ket{-1001001111}+ \nonumber \\ & a_{1}b_{3}\ket{3-101011100}+a_{1}b_{3}\ket{2-101011101}+a_{1}b_{3}\ket{0-101011110}+a_{1}b_{3}\ket{-1-101011111}+ \nonumber \\ & 
a_{1}b_{3}\ket{3-301101100}+a_{1}b_{3}\ket{2-301101101}+a_{1}b_{3}\ket{0-301101110}+a_{1}b_{3}\ket{-1-301101111}+ \nonumber \\ & 
a_{1}b_{3}\ket{3-401111100}+a_{1}b_{3}\ket{2-401111101}+a_{1}b_{3}\ket{0-401111110}+a_{1}b_{3}\ket{-1-401111111}+ \nonumber \\ &
a_{2}b_{0}\ket{1410000000}+a_{2}b_{0}\ket{0410000001}+a_{2}b_{0}\ket{-2410000010}+a_{2}b_{0}\ket{-3410000011}+\nonumber  \\ & a_{2}b_{0}\ket{1310010000}+a_{2}b_{0}\ket{0310010001}+a_{2}b_{0}\ket{-2310010010}+a_{2}b_{0}\ket{-3310010011}+ \nonumber \\ & a_{2}b_{0}\ket{1110100000}+a_{2}b_{0}\ket{0110100001}+a_{2}b_{0}\ket{-2110100010}+a_{2}b_{0}\ket{-3110100011}+ \nonumber \\ & a_{2}b_{0}\ket{1010110000}+a_{2}b_{0}\ket{0010110001}+a_{2}b_{0}\ket{-2010110010}+a_{2}b_{0}\ket{-3010110011}+ \nonumber \\ & 
a_{2}b_{1}\ket{1310000100}+a_{2}b_{1}\ket{0310000101}+a_{2}b_{1}\ket{-2310000110}+a_{2}b_{1}\ket{-3310000111}+ \nonumber \\ & a_{2}b_{1}\ket{1210010100}+a_{2}b_{1}\ket{0210010101}+a_{2}b_{1}\ket{-2210010110}+a_{2}b_{1}\ket{-3210010111}+ \nonumber \\ & a_{2}b_{1}\ket{1010100100}+a_{2}b_{1}\ket{0010100101}+a_{2}b_{1}\ket{-2010100110}+a_{2}b_{1}\ket{-3010100111}+ \nonumber \\ & a_{2}b_{1}\ket{1-110110100}+a_{2}b_{1}\ket{0-110110101}+a_{2}b_{1}\ket{-2-110110110}+a_{2}b_{1}\ket{-3-110110111}+ \nonumber \\ &
a_{2}b_{2}\ket{1110001000}+a_{2}b_{2}\ket{0110001001}+a_{2}b_{2}\ket{-2110001010}+a_{2}b_{2}\ket{-3110001011}+ \nonumber \\ & a_{2}b_{2}\ket{1010011000}+a_{2}b_{2}\ket{0010011001}+a_{2}b_{2}\ket{-2010011010}+a_{2}b_{2}\ket{-3010011011}+ \nonumber \\ & a_{2}b_{2}\ket{1-210101000}+a_{2}b_{2}\ket{0-210101001}+a_{2}b_{2}\ket{-2-210101010}+a_{2}b_{2}\ket{-3-210101011}+ \nonumber \\ & 
a_{2}b_{2}\ket{1-310111000}+a_{2}b_{2}\ket{0-310111001}+a_{2}b_{2}\ket{-2-310111010}+a_{2}b_{2}\ket{-3-310111011}+ \nonumber \\ & 
a_{2}b_{3}\ket{1010001100}+a_{2}b_{3}\ket{0010001101}+a_{2}b_{3}\ket{-2010001110}+a_{2}b_{3}\ket{-3010001111}+ \nonumber \\ & a_{2}b_{3}\ket{1-110011100}+a_{2}b_{3}\ket{0-110011101}+a_{2}b_{3}\ket{-2-110011110}+a_{2}b_{3}\ket{-3-110011111}+ \nonumber \\ & 
a_{2}b_{3}\ket{1-310101100}+a_{2}b_{3}\ket{0-310101101}+a_{2}b_{3}\ket{-2-310101110}+a_{2}b_{3}\ket{-3-310101111}+ \nonumber \\ & 
a_{2}b_{3}\ket{1-410111100}+a_{2}b_{3}\ket{0-410111101}+a_{2}b_{3}\ket{-2-410111110}+a_{2}b_{3}\ket{-3-410111111}+ \nonumber \\ &
a_{3}b_{0}\ket{0411000000}+a_{3}b_{0}\ket{-1411000001}+a_{3}b_{0}\ket{-3411000010}+a_{3}b_{0}\ket{-4411000011}+ \nonumber \\ & a_{3}b_{0}\ket{0311010000}+a_{3}b_{0}\ket{-1311010001}+a_{3}b_{0}\ket{-3311010010}+a_{3}b_{0}\ket{-4311010011}+ \nonumber \\ & a_{3}b_{0}\ket{0111100000}+a_{3}b_{0}\ket{-1111100001}+a_{3}b_{0}\ket{-3111100010}+a_{3}b_{0}\ket{-4111100011}+ \nonumber \\ & a_{3}b_{0}\ket{0011110000}+a_{3}b_{0}\ket{-1011110001}+a_{3}b_{0}\ket{-3011110010}+a_{3}b_{0}\ket{-4011110011}+ \nonumber \\ &
a_{3}b_{1}\ket{0311000100}+a_{3}b_{1}\ket{-1311000101}+a_{3}b_{1}\ket{-3311000110}+a_{3}b_{1}\ket{-4311000111}+ \nonumber \\ & a_{3}b_{1}\ket{0211010100}+a_{3}b_{1}\ket{-1211010101}+a_{3}b_{1}\ket{-3211010110}+a_{3}b_{1}\ket{-4211010111}+ \nonumber \\ & a_{3}b_{1}\ket{0011100100}+a_{3}b_{1}\ket{-1011100101}+a_{3}b_{1}\ket{-3011100110}+a_{3}b_{1}\ket{-4011100111}+ \nonumber \\  & a_{3}b_{1}\ket{0-111110100}+a_{3}b_{1}\ket{-1-111110101}+a_{3}b_{1}\ket{-3-111110110}+a_{3}b_{1}\ket{-4-111110111}+ \nonumber \\ & 
a_{3}b_{2}\ket{0111001000}+a_{3}b_{2}\ket{-1111001001}+a_{3}b_{2}\ket{-3111001010}+a_{3}b_{2}\ket{-4111001011}+ \nonumber \\ & a_{3}b_{2}\ket{0011011000}+a_{3}b_{2}\ket{-1011011001}+a_{3}b_{2}\ket{-3011011010}+a_{3}b_{2}\ket{-4011011011}+ \nonumber \\ & a_{3}b_{2}\ket{0-211101000}+a_{3}b_{2}\ket{-1-211101001}+a_{3}b_{2}\ket{-3-211101010}+a_{3}b_{2}\ket{-4-211101011}+ \nonumber \\ & a_{3}b_{2}\ket{0-311111000}+a_{3}b_{2}\ket{-1-311111001}+a_{3}b_{2}\ket{-3-311111010}+a_{3}b_{2}\ket{-4-311111011}+ \nonumber \\ &
a_{3}b_{3}\ket{0011001100}+a_{3}b_{3}\ket{-1011001101}+a_{3}b_{3}\ket{-3011001110}+a_{3}b_{3}\ket{-4011001111}+ \nonumber \\ & a_{3}b_{3}\ket{0-111011100}+a_{3}b_{3}\ket{-1-111011101}+a_{3}b_{3}\ket{-3-111011110}+a_{3}b_{3}\ket{-4-111011111}+ \nonumber \\ & a_{3}b_{3}\ket{0-311101100}+a_{3}b_{3}\ket{-1-311101101}+a_{3}b_{3}\ket{-3-311101110}+a_{3}b_{3}\ket{-4-311101111}+ \nonumber \\ & a_{3}b_{3}\ket{0-411111100}+a_{3}b_{3}\ket{-1-411111101}+a_{3}b_{3}\ket{-3-411111110}+a_{3}b_{3}\ket{-4-411111111}).
 \label{Eqn4final}
\end{align}
On the above state, measurements can be performed in the respective bases, and unitary operators can reproduce the two-qubit states in bidirectional teleportation.
\bibliographystyle{unsrt}
\bibliography{ref.bib}
\end{document}